\documentclass[draft]{agujournal2019}
\usepackage{apacite}
\usepackage{url} %this package should fix any errors with URLs in refs.
\usepackage{graphicx}

\draftfalse
\journalname{Earth and Space Science}

\begin{document}

\title{Power-law size distributions in geoscience revisited}

\authors{\'Alvaro Corral\affil{1,2,3,4}
and \'Alvaro Gonz\'alez\affil{1}}

\affiliation{1}{
Centre de Recerca Matem\`atica,
Edifici C, Campus Bellaterra,
E-08193 Barcelona, Spain}
\affiliation{2}
{Departament de Matem\`atiques,
Universitat Aut\`onoma de Barcelona,
E-08193 Barcelona, Spain}
\affiliation{3}
{Barcelona Graduate School of Mathematics,
Edifici C, Campus Bellaterra,
E-08193 Barcelona, Spain}
\affiliation{4}{
Complexity Science Hub Vienna,
Josefst\"adter Stra$\beta$e 39,
1080 Vienna,
Austria}

\correspondingauthor{\'Alvaro Corral}{alvaro.corral@uab.es}

\begin{keypoints}
\item We reanalyze the size distribution of earthquakes, karst sinkholes, wildfires, tropical cyclones, rainfall clusters, and impact fireballs.
\item A same, objective, statistical fitting method is improved and applied, allowing faithful comparisons between data sets.
\item The method automatically identifies the truncations of power-law distributions, crossovers between power-law regimes, and the best fit in the comparison between power-law and log-normal tails.
\end{keypoints}

\begin{abstract}

The size or energy of diverse structures or phenomena in geoscience appears to follow power-law distributions. A rigorous statistical analysis of such observations is tricky, though. Observables can span several orders of magnitude, but the range for which the power law may be valid is typically truncated, usually because the smallest events are too tiny to be detected and the largest ones are limited by the system size.

We revisit several examples of proposed power-law distributions dealing with potentially damaging natural phenomena. Adequate fits of the distributions of sizes are especially important in these cases, given that they may be used to assess long-term hazard. After reviewing the theoretical background for power-law distributions, we improve an objective statistical fitting method and apply it to diverse data sets. The method is described in full detail and it is easy to implement.

Our analysis elucidates the range of validity of the power-law fit and the corresponding exponent, and whether a power-law tail is improved by a truncated log-normal. We confirm that impact fireballs and Californian earthquakes show untruncated power-law behavior, whereas global earthquakes follow a double power law.
Rain precipitation over space and time
and tropical cyclones
show a truncated power-law regime. Karst sinkholes and wildfires, in contrast, are better described by truncated log-normals, although wildfires also may show power-law regimes. Our conclusions only apply to the analyzed data sets, but show the potential of applying this robust statistical technique in the future.
\end{abstract}

\section{Introduction}

Power-law distributions, or more correctly, power-law-like probability distributions,
first appeared in the study of the natural world
in relation with some ``human affairs''.
It seems that the pioneer work was that of Vilfredo Pareto,
who, at the end of the 19th century,
reported one of such distributions
accounting for the wealth of individuals \cite{Pareto_book,Kagan_book}.
Some years later, Auerbach and Estoup
showed that the population of cities
and the frequency of words in texts, respectively, follow
essentially the same statistical pattern
\cite{Newman2004a}.
Since then, many
social \cite{Axtell,Clauset},
technological \cite{Adamic_Huberman},
communication \cite{Serra_scirep,Corral_Boleda,Moreno_Sanchez},
and biological systems \cite{Furusawa2003,Camacho_sole,Pueyo}
have been found to display what is now called Zipf's law \cite{Li02},
a type of power-law-like distribution
that appears when counting the number of entities
that constitute collections of entities,
with the remarkable characteristic that %in most cases
the power-law exponent takes values close to 2.

In geoscience, considerable interest in power-law distributions
appeared in the 80's of the past century.
It was the appealing work of Mandelbrot on fractals
\cite{Mandelbrot}
which drew attention to the distribution of sizes of diverse geological objects
and structures,
like lakes, faults, fault gouge,
oil reservoirs, sedimentary layers, and even asteroids \cite{Turcottebook}.
Nevertheless, some of these systems had been explored much earlier;
for instance,
Bennet studied fragments of broken coal in 1936,
Korcak dealt with the distribution of islands in 1940,
and the distribution of lunar craters was reported %in 1970.
in the 1930's separately by McDonald and Young \cite{Cross}.
\citeA{Turcottebook} provides valuable bibliography on these issues.
%CITAR hartmann 1969???
Remarkably, for these systems,
when size is measured in terms of a linear dimension (e.g. diameter),
the power-law exponents are considerably larger than 2; they
typically range from 2.4 to more than 4.

Around 1990,
after the illuminating theory of Bak and co-workers
on self-organized-critical phenomena \cite{Bak_book,Watkins_25years},
the interest in power-law distributions was reinforced,
in particular regarding the ``severity'' or ``size'' of natural disasters
and other catastrophic geophysical phenomena
that can be considered to happen in terms of ``avalanches'' (events which suddenly release energy slowly accumulated in the system over a period).
But it was 60 years earlier (1932)
that Wadati had assumed a power-law distribution
for the energy of earthquakes,
whereas Ishimoto and Iida published in 1939 a power law for the distribution
of earthquake amplitudes recorded by a microseismograph
\cite{Utsu_GR}.
Nowadays, these two power laws are known to be different representations of the
Gutenberg-Richter law of earthquake ``size'' \cite{Kagan_gji02,Kanamori_rpp}.
Following the seminal work of Bak \cite{BTW87,Bak_book},
many authors claimed power-law distributions
in systems as diverse as
volcanic eruptions,
rockfalls,
land slides,
forest fires,
solar flares,
pulsar glitches,
or biological extinctions \cite{Bak_book,Malamud_hazards}.
Again in contrast to Zipf's law,
the power-law exponent is not constrained to be around 2,
but shows considerable scatter between 1 and 2,
depending on the system under consideration.

An important drawback of these studies has been statistical rigor.
In most cases, the evidence for power-law distributions was just an apparent
linear behavior in a log-log representation of the probability density
or of the complementary cumulative distribution.
In some other cases, a linear-regression straight line was fit by
the least squares method, a procedure which can lead to substantial biases and wrong inferences
when applied to probability distributions \cite{White,Clauset}.
More recent works have made an effort to improve the statistical methodology.

In this paper we attempt to give an overview of power-law distributions in geoscience;
however, due to the heterogeneity of the approaches,
a proper and fair comparison of results is impossible,
and we have opted for a revision of paradigmatic systems
using a rigorous statistical protocol,
which is an improvement of a previous one \cite{Corral_Deluca}.
Although the definitive recipe to fit power-law distributions does not yet exist,
the method developed here is reasonable enough,
and fully objective.
Due to many limitations, our overview of power laws is far from systematic
and we have selected instead a few representative examples
across the geosciences.

We will concentrate on the (spatial) size of structures on the solid Earth and atmosphere
(karst sinkholes and rain clusters)
and on the size, in terms of energy or severity, of diverse natural events
(earthquakes, wildfires, tropical cyclones, amount of rainfall, and impact fireballs).
All these examples relate to potentially damaging natural phenomena,
for which adequate fits of their size distribution
are especially important,
given that these fits
(together with the temporal occurrence rate)
are used for assessing the long-term hazard these phenomena pose.
Further examples with potential power-law distributions
have been quoted in geology
\cite{Turcottebook,Burroughs_Tebbens},
hydrology \cite{Aban},
ecology \cite{White}, and
astrophysics \cite{Aschwanden_open}.
We will not be able to approach other interesting geophysical variables,
such as distances between events or jumps
\cite{Davidsen_distance,Corral_prl.2006,Felzer}
or
such as durations and waiting times \cite{Corral_csf},
which have also been related in some way to power-law distributions,
for example for
earthquakes \cite{Bak.2002,Corral_physA.2004},
volcanic eruptions \cite{Cannavo2016},
solar flares \cite{Boffetta,Baiesi_flares},
solar wind \cite{Wanliss}
or
geomagnetic storms \cite{Morina_storms}.
Possible power-law distributions of so-called intensive variables,
such as rain rate \cite{Yano_oneoverf,Peters_Deluca}
are also disregarded in this paper.

As mentioned in the first paragraphs, power-law-like distributions
are far from being exclusive of geoscience \cite{Li02}.
The interested reader can find extensive bibliography
for biological systems \cite{munoz_colloquium},
neuroscience \cite{Chialvo_review},
economy \cite{Farmer_power_laws},
and technology \cite{Adamic_Huberman}.
The ubiquity of power-law-like distributions
has induced some authors to claim for a common origin for them
\cite{BTW87,Bak_book}, although
a large variety of alternative explanations has been proposed
\cite{Simon,Miller_monkey,Sornette_critical_book,Mitz,Newman2004a,Saichev_Sornette_Zipf,Simkin11,Tria,Corominas_dice,Ferrer_cancho_compression,Penland}.
Our statistical analysis will not allow us to enter into mechanistic and generative models
and the debate associated to them;
nevertheless, in the last section we provide a summary of these explanations.

Thus, in the next section we explain untruncated and truncated power-law distributions,
and the nuances that distinguish them from power-law-like distributions.
Then, we expose our procedure to fit power-law-like distributions
(which also applies to the log-normal and could be immediately extended to any other distribution).
We also explain a likelihood-ratio test to distinguish between power-law tails and log-normal tails.
Finally, Sec. 4 explains the data sets analyzed as well as the results obtained from them.
The conclusions are at the end.

\section{Power-law distributions and power-law-like distributions}

Let us recall that the probability density $f(x)$ of a continuous random variable is defined
as the probability that the random variable is between the values $x$ and $x+dx$,
divided by $dx$, where $dx$ is the width of the interval (also called bin).
In the ideal mathematical case $dx$ goes to zero
\cite{Ross_firstcourse8}
but in practice $dx$ has to be wide enough
to make it possible the counting of several events per interval.
In geoscience $x$ may represent the size
of some geological structure
(faults, islands, lakes...) or
of some geophysical phenomenon
(energy of earthquakes, amount of rain, etc.).
It is noteworthy that $f(x)$ is not a probability,
but a density of probability, so it is a physical quantity with units of $x^{-1}$.

Equivalently, the probability distribution can be described in terms of the complementary
cumulative distribution function, $S(x)$, which provides the probability that the random variable
takes a value above $x$.
Both functions are related by means of % the relations
$$
f(x) = - \frac{d S(x)} {dx}
\mbox{ and }
S(x) = \int_x^\infty f(x') dx'.
$$
Note that $f(x)$ fulfills $f(x)\ge 0$ and $\int_{-\infty}^\infty f(x) dx=1$
(normalization),
whereas $S(x)$ is a non-increasing function of $x$
with $\lim_{x \rightarrow -\infty}S(x) =1$ (normalization) and
$\lim_{x \rightarrow \infty}S(x) =0$.
In some literature it is difficult to guess if the authors are dealing with
$f(x)$ or $S(x)$, or with some variation of any of them.
Sometimes, when $f(x)$ or $S(x)$ are estimated from data,
their dependence on $x$ is referred to as the
frequency-size relationship or even the size-frequency distribution.

\subsection{Untruncated power-law distribution}

A % quantity
continuous
variable $x$ is power-law (pl) distributed
\cite{Johnson_univariate_continuous}
if its probability density is given by
$$
f_{pl}(x)=\frac C {x^\beta}
$$
for $x \ge a$,
and $f_{pl}(x)=0$  for $x < a$;
with $\beta > 1$
and $a$ the lower cut-off
or lower truncation, fulfilling $a>0$.
Normalization implies that the normalization constant $C$
is determined by $a$ and $\beta$, and so
\begin{equation}
f_{pl}(x)=\frac{\beta-1}{a}\left(\frac{a}{x}\right)^{\beta}
\label{powerlawf}
\end{equation}
for $x\ge a$.

In terms of the complementary cumulative distribution the power law is defined as
\begin{equation}
S_{pl}(x)= \left(\frac{a}{x}\right)^{\beta-1}
\label{powerlawS}
\end{equation}
for $x\ge a$ and $S_{pl}(x)=1$  for $x < a$.
Note that there is some ambiguity regarding the power-law exponent,
which is $\beta$ for $f_{pl}(x)$ and $\beta-1$ for $S_{pl}(x)$.
We take the convention that the exponent of the distribution is $\beta$,
the exponent of the density.
In order to distinguish the power-law distribution from the
(upper)
truncated power law, introduced below,
sometimes we may talk about ``untruncated power law'', referring to the former,
although both distributions are unavoidably truncated from below ($a>0$).

Two important properties of power-law distributions are scale invariance
and divergence of moments.
The first one means that power laws remain the same after appropriate
changes of scale in $x$ and $f(x)$; for instance if $f(x) \propto 1/x^{3/2}$
the change $x\rightarrow 100 x$ and $f\rightarrow f/1000$ leaves the resulting
function the same
(this does not happen with an exponential, for instance,
nor with any other function different than the power law).
Thus, scale invariance implies that no characteristic scale exists
\cite{Takayasu,Christensen_Moloney,Corral_Lacidogna}.
Nevertheless, strictly speaking,
scale invariance holds for power laws defined in the whole range $x>0$
(for instance, Newton's law of gravitation),
whereas, due to normalization,
power-law distributions are defined for $x\ge a > 0$,
this prevents true scale invariance;
in other words, the value of $a$ sets a characteristic scale,
so one only may talk of scale invariance above the lower cut-off $a$.

Interestingly, for some systems the cut-off $a$ seems to be so small
that its value is unknown, which implies that no characteristic scale shows up.
This is the case of earthquakes, for instance.
So, a question about ``which the typical size of earthquakes in Japan or
California is'', cannot be answered, as no characteristic size can be defined.
Due to their scale-invariant properties,
power-law distributions are sometimes called fractal distributions, as by \citeA{Turcottebook};
nevertheless, we prefer the former terminology,
which does not contain any implicit reminiscence of self-similar geometry
(one can ignore spatial structure, if it exists).

Divergence of moments means that if, for instance,
the exponent is $\beta \le 2$,
the mean or expected value of the variable (the first moment),
$\langle x \rangle = \int_{-\infty}^\infty x f(x) dx$,
becomes infinite, in the same way that all higher moments,
$\langle x^k \rangle = \int_{-\infty}^\infty x^k f(x) dx$,
with $k \ge 1$.
If $2 < \beta \le 3$ the first moment is finite but
the second moment $\langle x^2\rangle$ and higher moments
become infinite, and so on.
Divergence of moments has the annoying consequence that some
important results of probability theory do not hold,
as the law of large numbers when $\beta \le 2$ \cite{Shiryaev,Corral_csf}.
So, in this case
one cannot estimate the mean of the distribution from the sample mean,
simply because the mean of the distribution is infinite and the sample mean
cannot converge to this value \cite{Corral_FontClos}.
If $2 < \beta \le 3$ the standard central-limit theorem does not apply
and the sample mean neither follows Gaussian statistics
nor has a finite variability
\cite{Bouchaud_Georges}.

If one takes logarithms on the expressions of $f_{pl}(x)$ or $S_{pl}(x)$
for a power-law distribution,
one immediately realizes that both $\ln f_{pl}(x) $ and $\ln S_{pl}(x)$
are linear functions of $\ln x$.
If the least-squares method is applied to the logarithm
of the empirical estimations of $f(x)$ or $S(x)$
one can get an estimation of the parameters $\beta$ and $a$.
However, one cannot do inference with the resulting parameters,
as some requirements of the theory of linear regression
are not fulfilled.
In addition, obtaining an accurate empirical estimation of $f(x)$ is not straightforward,
as one needs to choose bins in a proper way \cite{Wand,Corral_Deluca}.
Let us clarify that the least-squares method is inappropriate to estimate probability distributions,
but not to establish linear correlation between two variables, $x$ and $y$
(or $\log x$ and $\log y$), for instance,
nor to curve fitting in general.

Despite the peculiar properties of power-law distributions,
estimation of their parameters is straightforward
using maximum likelihood.
This is a method of estimation of parameters that
gathers a number of desirable properties \cite{casella},
such as
invariance under re-parameterization
(the resulting estimated distribution does not depend on the choice of parameters)
and invariance under change of variables
(the resulting description of the phenomenon does not depend on the
selected random variable).
When applied to probability
distributions that belong to the so-called exponential family
(both power laws and truncated power laws belong to this family,
as well as the log-normal, truncated or not),
maximum-likelihood estimators
turn out to fulfill consistency
(they are unbiased and convergent, asymptotically;
that is, the estimation tends to the true value if the number of data is large),
due to the uniqueness of the maximum \cite{pawitan2001},
and also fulfill asymptotic efficiency
(they achieve the Cram\'er-Rao lower bound asymptotically,
i.e., the estimation tends to have the smallest possible variance among all the unbiased estimators)
under some standard regularity conditions \cite{pawitan2001}.

The maximum-likelihood estimation of the power-law exponent yields
\cite{Clauset,Corral_Deluca}
\begin{equation}
\beta_{pl}= 1+ \frac 1 {\ln (g/a)},
\label{mle_pl}
\end{equation}
where $g$ is the geometric mean of the sample,
defined as $\ln g =n^{-1}\sum_{i=1}^n \ln x_i$,
with $n$ the size of the sample and
$x_i$ the $n$ observations of the random variable.
As $\beta$ is determined from the sample,
it is subject to statistical uncertainty (it is a ``statistic'')
and its standard deviation, or standard error is
\begin{equation}
\sigma_\beta=\frac {\beta-1}{\sqrt n},
\label{sigmabeta}
\end{equation}
where $\beta$ has to be understood here as the true value of the exponent
(not the estimated one, although they will be very close if $n$ is large).
Regarding the cut-off $a$, it is estimated as
$a_{pl}=\min(x_1,\dots x_n)$.
The reason is that, as long as $x_i \ge a$ for all $i$,
the larger $a$, the higher the likelihood.
However, if a single $i$ fulfills $x_i < a$ then the likelihood
becomes zero (as $f_{pl}(x_i)=0$).
This implies that $a$ has to grow up to the smallest
value of $x_i$, but not more.

Which is then the problem with power laws, if this estimation procedure is so straightforward?
The key issue is that in practice one does not deal with (pure) power-law distributions,
but with power-law-like distributions.
These constitute a loose family of distributions
\cite{Farmer_power_laws}
with the characteristic that,
for a certain range of $x$, the distribution resembles in some undefined way a power law.
Consider the so-called full-tails gamma (ftg) distribution
\cite{ftg}
$$
f_{ftg}(x)=\frac 1 {\theta_2 \Gamma(1-\beta,\theta_1/\theta_2)}
\left(\frac{\theta_2}{\theta_1+x}\right)^\beta
\exp\left(-\frac{\theta_1+x}{\theta_2}\right),
$$
with $\theta_1>0$, $\theta_2>0$, and $-\infty < \beta < \infty$
and with $\Gamma(1-\beta,\theta_1/\theta_2)$ the (upper) incomplete gamma function
\cite{Abramowitz}.
This distribution is a truncated gamma distribution
\cite{Serra_Corral}
extended to $1-\beta < 0$
and shifted to have support
in the interval $[0,\infty)$.
Let us consider $\theta_1 \ll \theta_2$.
In this case the distribution resembles a power law in some
range $a\le x \le b$, with $\theta_1 \ll a$ and $b \ll \theta_2$,
but, strictly speaking, that part of the distribution is not a power law.
Interestingly, recent work \cite{Voitalov_krioukov}
has identified power-law-like distributions
with regularly varying distributions, for which $S(x)$ is a power law multiplied by a
slowly varying (unknown) function.
This leads to asymptotic power laws, which is a concept in some cases similar but certainly different to our power-law-like behavior (for which $b$ can be finite).

In practice,
the disadvantage is that
the real underlying theoretical distribution $f(x)$ is unknown.
For instance, simple branching processes (bp)
(which are mean-field models by construction)
yield discrete probability distributions
for their total number of elements $x$,
which
(close to their critical point
and
for large $x$)
lead to truncated gamma tails,
$f_{bp}(x) \propto x^{-3/2} e^{-x/\theta}$ \cite{Corral_FontClos};
but branching processes with finite-size effects
lead to more complicated functional forms for the tail, even in the critical case
\cite{Corral_garcia_moloney_font}.
Beyond mean field little is known,
and of course, real systems are more complicated than any model.
On the other side, incompleteness effects for small values of $x$
may provoke an underestimation on the count of $x$
and a deviation from a power-law behavior, as
modeled for instance by the full-tails gamma distribution above, or by the
genuine Pareto (par) distribution
$f_{par}(x)=(\beta-1) \theta^{\beta}/(\theta+x)^\beta$;
nevertheless, this sort of modelling is ad-hoc.

In summary, there are a number of processes and factors that
trigger deviations from power-law behavior both for small and large values
of $x$; however, these factors are difficult to parameterize.
If one can disregard large$-x$ deviations
(perhaps because the number of data is so low that the scale of deviations
is not sampled)
one can model the data by a (pure) untruncated power-law distribution,
Eqs. (\ref{powerlawf})--(\ref{powerlawS}),
but the meaning of the parameter $a$ is totally different than in the original definition, as $a$ defines the scale at which the small$-x$ deviations become negligible,
and values of $x$ below $a$ need to be disregarded.
The rest of values of $x$, those above $a$,
will be considered in this paper to define the tail of the distribution.
The main problem in fitting power-law-like distributions is
finding the lower cut-off $a$.
Note that in this context
we will need to distinguish between the total number of data $N$
and the number of data in the power-law range, $n$.

\subsection{Truncated power-law distribution}

If, in addition,
there are deviations from power-law behavior for large values of $x$
and there is no information on the shape of the tail,
the situation gets worse.
An option is to fit a truncated power-law (tpl) distribution
\cite{Johnson_univariate_continuous,Aban,Burroughs_Tebbens,Corral_Deluca},
given by
$$
f_{tpl}(x)=\frac{\beta-1}{a\left(1-c^{\beta-1}\right)}
\left(\frac a x\right)^{\beta},
$$
for $a \le x \le b$
and $f_{tpl}(x)=0$ otherwise; with $a>0$, $b>0$, $c=a/b$, and $-\infty < \beta < \infty$
(nevertheless, the case of negative exponent corresponds to an increasing
power law and is of little interest;
the case $\beta<1$ allows $a\ge 0$).
The limiting case $\beta=1$ needs a separate formulation \cite{Corral_Deluca}.
It has to become clear that we apply this distribution to the central part of the data,
disregarding values of $x$ outside the interval $[a,b]$.
Of course, the major problem is to find appropriate values of $a$ and $b$.
Again, we will distinguish between the total number of data $N$
and the number of data in the power-law range, $a \le x \le b$, denoted as $n$.

The alternative description in terms of the complementary cumulative distribution yields
$$
S_{tpl}(x)=\frac {(a/x)^{\beta-1} - c^{\beta-1}}
               {1 - c^{\beta-1}}
$$
for $a \le x \le b$;
$S_{tpl}(x)=1$ for $x<a$ and
$S_{tpl}(x)=0$ for $x>b$.
The limit $b\rightarrow \infty$ when $\beta > 1$
returns to the usual power law,
Eqs. (\ref{powerlawf})--(\ref{powerlawS}).
Note that for the truncated power law,
$\ln f_{tpl}(x)$ is still a linear function of $\ln x$,
but $\ln S_{tpl}(x)$ is not.
When $\beta > 1$ linearity between $\ln S_{tpl}(x)$ and $\ln x$
only takes place for $x\ll b$.

The maximum-likelihood estimation of the exponent $\beta$
cannot be solved explicitly
and one is faced to the numerical maximization of the likelihood,
or, equivalently, of the per-datum log-likelihood
\cite{Corral_Deluca}
\begin{equation}
\mathcal{L}_{tpl}(\beta) = \ln \frac{\beta-1}{1-c^{\beta-1}}
- \beta \ln \frac g a -
\ln a.
\label{mle_tpl}
\end{equation}
%with $c=a/b$
Care with numerical overflows
must be taken when $\beta$ gets close to one,
see \citeA{Corral_Deluca}.
An analytical expression for the standard deviation of $\beta$
can be derived \cite{Aban,Corral_Deluca};
nevertheless, we will not use it (the reasons will become clear later, Sec. \ref{subsec32}).
In the same way, the maximum-likelihood estimators of $a$ and $b$,
which are $a_{tpl}=\min\{x_1,\dots x_n\}$ and
$b_{tpl}=\max\{x_1,\dots x_n\}$,
will be meaningless for us, as these estimations assume that the data set is fixed,
but we need to find precisely which subset of the data may follow a (truncated) power-law.

\subsection{Double power-law distribution}

It may happen that some data set is well fit by a truncated power law from $x=a$
up to some value $x=b$, and from $x>b$ the data is also well fit by an untruncated power law.
Let us relabel this crossover point as $b=\theta$.
Then, it is clear that the data is fit by two power-law regimes for $x>a$,
and we may define the double power-law (dpl) distribution as
$$
f_{dpl}(x)=(1-q) \frac{\beta_1-1}{\theta} \frac 1 {c^{1-\beta_1}-1} \left(\frac \theta x \right)^{\beta_1} \mbox{ for } a\le x \le \theta,
$$
$$
f_{dpl}(x)=     q \frac{\beta_2-1}{\theta}                                     \left(\frac \theta x \right)^{\beta_2} \mbox{ for } \theta \le x,
$$
and zero for $x<a$,
and where the parameter $q$ is chosen to ensure continuity between the two regimes at $x=\theta$,
leading to
$$
q=\frac{\beta_1-1} {(\beta_2-1) c ^{1-\beta_1} - (\beta_2-\beta_1)},
$$
and the two exponents fulfilling $-\infty < \beta_1 < \infty$ and $\beta_2> 1$,
$c=a/\theta$, and
with $a >0$     if $\beta_1 > 1$
and  $a \ge 0$ if $\beta_1 < 1$.
The complementary cumulative distribution function leads to
$$
S_{dpl}(x)=q+(1-q)\frac {(\theta/x)^{\beta_1-1} - 1 }{c^{1-\beta_1}-1 }  \mbox{ for } a\le x \le \theta,
$$$$
S_{dpl}(x)= q \left(\frac \theta x\right)^{\beta_2-1} \mbox{ for } \theta \le x,
$$
and zero for $x<a$.
The sudden change of slope at $x=\theta$ is unlikely to fit real-world distributions with
a large number of data, so some refinement in the parameterization may be needed
to fit properly the crossover in this case.

\subsection{Change of variables}

A noteworthy issue comes from the fact that the variable $x$ is not
uniquely defined.
For example, in terms of size of objects, $x$ can be the linear size,
denoted by $L$ or can be the volume $V$.
Assuming a general relation between both, $V \propto L^z$
(with $z=3$ for spherical or cubic objects),
it can be shown that if one of them is power-law distributed,
so it is the other,
both for untruncated and for truncated power laws
(and also for the double power law),
and the power-law exponents are related by
\begin{equation}
z=\frac{\beta_L-1}{\beta_V-1}.
\label{zetabeta}
\end{equation}
This relation explains, at least in part, why the exponents $\beta_L$ of the distributions
of geological objects become so large (even larger than 4)
when the size is defined in terms of the linear size $L$.
In terms of the volume $V$ the exponents become smaller.
Note that this simple relation does not apply to
the more complicated scenario that arises, for example, when volumes
have to be calculated from random cross sections \cite{GAONACH_lovejoy};
see also \citeA{Stereology}.

\section{Fitting of power-law-like distributions}

\subsection{The Clauset \textit{et al.}'s procedure}

Drawbacks in fitting power-law distributions were pointed out by \citeA{Clauset},
but were previously noticed by other authors, such as \citeA{Goldstein,Bauke,White}.
There is an important confusion regarding the method and achievements of \citeA{Clauset},
as many authors claim that they are using this method while they are just fitting
by plain maximum-likelihood estimation.
Summarizing, the Clauset \textit{et al.}'s method
(for an untruncated power-law distribution)
proceeds in two parts.
For the first part one calculates a tentative fit as follows
\begin{itemize}
\item
Pick a value of the lower cut-off $a$.
\item Fit, by maximum likelihood, a power law to the range $x\ge a$
(using Eq. (\ref{mle_pl})). A value of $\beta$ is obtained.
\item Calculate the Kolmogorov-Smirnov distance between
the empirical distribution (for $x\ge a$) and the theoretical distribution
(using the value of $\beta$ obtained in the previous step).
\end{itemize}
Repeat the procedure for all possible values of the lower cut-off,
and select the one which yields the minimum Kolmogorov-Smirnov distance.
The selected value of $a$ has associated a value of the exponent $\beta$,
and both parameters define the tentative fit.

The second part of the procedure assigns a $p-$value to the fit.
One only needs to generate samples
with the same number of data than the original empirical data.
These samples
are obtained from bootstrap of the empirical data for $x<a$ and
are simulated synthetic power laws
for $x\ge a$ (with the resulting value of $\beta$).
Applying the first part of the procedure to any of the synthetic samples,
one obtains a distribution of the minimized Kolmogorov-Smirnov distance,
which allows one to define a $p-$value as the probability that
the minimized distance takes a value
larger than the one obtained empirically.
At the end, the recipe is:
Reject the tentative fit if the $p-$value is too low,
according to the desired significance level;
otherwise, there is no reason for rejection and the power-law fit is ``accepted''.

Unfortunately, we have encountered several drawbacks of the Clauset \textit{et al.}'s method.
First, the method is ad-hoc and there is no justification why the minimization
of the Kolmogorov-Smirnov distance should work to find a meaningful value of $a$
\cite{Corral_Deluca}.
Second, the method cannot be extended to truncated power-law distributions
(i.e., it is only prescribed for $b\rightarrow \infty$).
And third, we have found the method to fail when
applied to simulated data with real power-law tails, since the power law is rejected, despite it is a real, synthetic power law \cite{Corral_nuclear}.
Other authors have criticized also Clauset \textit{et al.}'s method
\cite{Voitalov_krioukov}.
In general, power-law fitting is a controversial issue
\cite{Barabasi_criticism,Holme_scalefree}.
Therefore, despite the popularity of the Clauset \textit{et al.}'s method, we have developed
an alternative procedure.

\subsection{Alternative power-law fitting procedure \label{subsec32}}

Our method is close somehow to the one of Clauset \textit{et al.},
in the sense that
it
is based on both maximum likelihood estimation \cite{pawitan2001} and on the
Kolmogorov-Smirnov goodness-of-fit test \cite{Press}.
The version presented here is a straightforward extension of a previous work
\cite{Peters_Deluca,Corral_Deluca}.
We explain the case of truncated power laws,
simplification to untruncated power laws is trivial:

\begin{itemize}
\item
Pick a value of the lower cut-off $a$
and another value of the upper cut-off $b$.
\item Fit, by maximum likelihood, a truncated power law to the range $a\le x\le b$
(maximizing Eq. (\ref{mle_tpl})). A value of $\beta$ is obtained.
\item Calculate the Kolmogorov-Smirnov distance between
the empirical distribution (restricted to $a \le x\le b$) and the theoretical distribution
(using the value of $\beta$ obtained above).
\item Assign a $p-$value to the resulting tentative fit,
in the following way:
\begin{itemize}
\item
Simulate a synthetic power law in the range $a \le x \le b$,
with the value of $\beta$ just obtained,
and with the same number of data $n$.
\item
Apply to the synthetic data the previous two steps
(fit of $\beta$ and calculation of Kolmogorov-Smirnov distance
with the new value of $\beta$).
\end{itemize}
%Simulate new synthetic power laws
Repeat the simulation procedure many times in order to obtain the distribution
of the Kolmogorov-Smirnov distance
(the theoretical distribution cannot be used because $\beta$ is calculated
from the same data to which we apply the test).
The $p-$value is obtained as the probability that this distance is larger than
the empirical distance.
%\end{itemize}
\end{itemize}
Repeat this procedure for all possible values of $a$ and $b$.
If more than one pair of values yield ``acceptable'' $p-$ values
(higher than the significance level $p_{min}$),
choose the pair of $a$ and $b$ that leads to the largest ratio $b/a$
(this is the largest power-law log-range), and the resulting values of
$a$, $b$, and $\beta$ give the resulting ``accepted'' fit.
If no high enough $p-$values are found, the power law is rejected.
In this paper we take a demanding significance level $p_{min}=0.20$.
Note that this applies to continuous random variables, while
the fitting of discrete power laws is a bit more involving \cite{Corral_Deluca_arxiv}.

This is the original procedure \cite{Peters_Deluca, Corral_Deluca},
which has an important drawback:
it does not allow to estimate the uncertainty of the cut-offs $a$ and $b$.
Moreover, the uncertainty in the exponent $\beta$ was obtained
from the standard deviation of the maximum-likelihood estimation,
Eq. (\ref{sigmabeta}) or the equivalent one for the truncated case \cite{Aban},
which assumed $a$ and $b$ fixed.
Therefore, we expect the real uncertainty in $\beta$ to be larger.

This problem, or these two related problems, are solved here in a
very simple way.
We
just take bootstrap resamplings of the original data
\cite{Good_resampling}
and repeat the whole procedure
with them; this will allow to obtain distributions for $a$, $b$, and $\beta$,
from which their uncertainty can be estimated.
This method will be applied to diverse geophysical data in the next sections.
Moreover, as the distributions of both cut-offs
are somewhat asymmetric,
we consider their logarithms (which are more symmetric)
and report the uncertainty of the cut-offs
using one standard deviation of the logarithms.
Bootstrap was used in a similar context by \citeA{Woessner2005},
in order to estimate the uncertainty of the magnitude of completeness
of seismic catalogs.
Note that, although it has been recognized that bootstrap
can lead to biased information when dealing with extreme-value distributions,
our estimation procedure does not make use of those distributions;
in fact, the only statistic that is involved in the estimation is a mean value
(that of the logarithm of the random variable, Eqs. (\ref{mle_pl}) and (\ref{mle_tpl})).

The method can be complemented by studying the dependence
of the exponent $\beta$ on both cut-offs \cite{Baro_Vives},
taking cut-off values inside the power-law range found $(a,b)$.
In a real power law, $\beta$ should be stable against increases in $a$ and decreases in $b$;
conversely, a trend of $\beta$ as a function of $a$ or $b$
is an indication of a spurious power law
(it may happen when the number of data is low,
as then rejection of the power law gets more difficult).
Interestingly, the resulting dependence on the cut-offs could be used to identify
an alternative distribution to the power law, if the resulting exponent does not show a flat behavior \cite{Salje_Vives}.

For the fitting of an untruncated power-law distribution we introduce an additional improvement.
We impose that the range of variation of the lower cut-off $a$ does not cover the whole range of $x$,
but only the range $a >a_{cv}$ for which an untruncated power law
can be considered to be a better fit than a truncated log-normal.
This is developed in the next subsection.

\subsection{Comparison of untruncated power law versus truncated log-normal}

In addition to evaluate if a power-law distribution is a good fit in some range of a particular data set,
one may compare the power law in front of diverse alternative distributions \cite{Clauset}.
Particularly important and widely used in the same context
\cite{Turcottebook,Corral_fires,Malevergne_Sornette_umpu,Hantson_Pueyo16},
is the log-normal case
\cite{Johnson_univariate_continuous,Limpert_lognormal},
for which a likelihood ratio test can be applied in a very simple way,
comparing the untruncated power law with the truncated log-normal
(the term truncated here refers to a lower truncation of the log-normal,
the power law is always lower truncated, due to normalization).
In the most general case,
the truncated log-normal (tln) distribution is defined by the probability density
$$
f_{tln}(x)=
{\sqrt{\frac 2\pi}}
\left[
 \mbox{erfc}\left(\frac{\ln a -\mu}{\sqrt{2} \sigma}\right)
-\mbox{erfc}\left(\frac{\ln b -\mu}{\sqrt{2} \sigma}\right)
\right]^{-1}
\frac 1{ \sigma x}
\exp\left(-\frac{(\ln x-\mu)^2}{2\sigma^2}\right),
$$
%for $x$ positive
for $a \le x \le b$
(and zero otherwise);
with $-\infty < \mu < \infty$,
$\sigma$ positive,
the cut-offs $a$ and $b$ also positive,
and $\mbox{erfc}(y)=\frac 2 {\sqrt{\pi}} \int_y^\infty e^{-x^2} dx$ the complementary error function
(unfortunately, with this parameterization $\mu$ and $\sigma$ have no clear physical meaning).
Except in one particular case, we consider $b^{-1}$ fixed to 0,
for which the second erfc term goes to zero.
This distribution will be sometimes simply referred here as log-normal,
for economy of language.
The true (untruncated) log-normal distribution is recovered in the limit $a\rightarrow 0$ and $b\rightarrow\infty$.
Note that, in contrast to other fitting methods, where the normalization constant can be an additional parameter,
both maximum-likelihood estimation and the Kolmogorov-Smirnov test require the exact computation
of the normalization constant (this is the reason of the apparent complication of the expression for $f_{tln}(x)$).

The comparison procedure is based in the calculation of the residual coefficient of variation
$cv_\ell$
of the logarithm of the rescaled random variable
\cite{Malevergne_Sornette_umpu}.
We need to define
$x_{(i)}$ as the $i-$th value of the variable when this is sorted
in an increasing way, i.e., $x_{(1)} \le x_{(2)} \le \dots  \le x_{(N)} $.
Then,
$$
cv_\ell=\frac{s_\ell}
{m_\ell},
$$
with $m_\ell$ the mean of the new variable $\ell$ defined as the logarithm of the variable rescaled by $x_{(k)}$ , i.e.,
$\ell=\log(x/x_{(k)})$,
which is calculated for the $n=N-k$ values of the original variable fulfilling $x > x_{(k)}$
(we assume the variable $x$ is continuous and disregard the case in which some value of $x$
is repeated in the data set);
so,
$$
m_\ell=\frac 1 n {\sum_{i=k+1}^N \log \frac {x_{(i)}}{x_{(k)}}}.
$$
In addition, $s_\ell^2$ is the unbiased variance of $\ell$, i.e.,
$$
s_\ell^2=\frac 1 {n-1} {\sum_{i=k+1}^N \left( \log \frac {x_{(i)}}{x_{(k)}} - m_\ell \right)^2}.
$$
The base of the logarithm does not matter,
as long as it is the same in the computation of $m_\ell$ and $s_\ell$.

The test is based on the idea that, for an untruncated power-law distribution,
the logarithmic coefficient of variation of a sample is close to 1
(it is exactly equal to 1 in theory).
The critical values of the test can be obtained simulating a power law distribution
(the exponent does not matter) and calculating the distribution of $cv_\ell$.
In practice, this is equivalent to calculating the usual coefficient of variation
(standard deviation divided by mean, without logarithms)
of an exponential distribution with unit scale parameter,
which is obtained from $-\ln(1-u)$,
with $u$ a uniform random number in $[0,1)$.
If the empirical value of the logarithmic coefficient of variation is between
the percentiles 5 and 95 of the simulated values,
the power-law null hypothesis cannot be rejected,
with a 90\% confidence.
If the logarithmic coefficient of variation is too small (below percentile 5)
the power law must be rejected in favor of the log-normal,
and if it is too high (above percentile 95) the power law is rejected,
but the alternative is not the log-normal
(this latter option does not happen in any of the cases analyzed here).

This simple test is the uniformly most powerful unbiased test
for comparing the two alternatives,
and is rooted on the facts that the power-law distribution is nested
into the truncated log-normal,
and that their likelihood ratio is a decreasing function of the
logarithmic coefficient of variation \cite{Castillo}.
The former fact means that the power law can be understood
as a special case of a truncated log-normal
(with its parameters $\mu$ and $\sigma$ fulfilling
$\mu -\ln a\rightarrow -\infty$,
$\sigma^2 \rightarrow \infty$,
and
$\beta=1+|\mu-\ln a|/\sigma^2$,
taking into account that $\ln (x/a)$ is a truncated normal),
and one evaluates whether the parameters of the log-normal
are significantly different from the power-law limit of the log-normal.
So, strictly speaking, one never rejects a log-normal,
but rejects a log-normal different than its power-law limit.
When performing the test for different values of $k$
in order to find a transition from log-normal to power law
(when increasing the cutting index $k$,
or the opposite transition when decreasing $k$),
the test becomes somehow ``subjective'' as sometimes the crossing
of the critical region is erratic.
In any case, this will provide us with a value of the cut-off $a$,
called now $a_{cv}$ in order to distinguish it from the one coming from the fit,
signaling the value above which a log-normal fit does not bring any improvement
with respect an untruncated power law.
In consequence, the range of variation of $a$ in the previous section will be $a> a_{cv}$
(for the untruncated power law only).
In some cases, although a power-law tail cannot be rejected
in front of the log-normal for some range,
the log-normal will provide a much larger fitting range
(beyond the tail) and will be therefore preferred.

Note that, in general, a likelihood ratio test does not tell us if any of the two fits is good or bad,
rather, it yields
which one of the two options has higher likelihood and if the difference is significant or not.
So, even if the power-law is rejected in front of the log-normal,
the latter can be a bad fit,
or, on the contrary, the power-law can be a good fit and the log-normal an even better one.
The likelihood ratio test presented here, based on the logarithmic coefficient of variation of the tail,
has the peculiarity that it does not need the calculation of the fitting parameters;
this is an advantage from the computational point of view, as its implementation is very simple,
but if one is interested in the parameters these have to be obtained separately.
Then, in order to obtain the truncated log-normal fits we proceed in a way totally analogous
as for the truncated power law (sweeping different values of $a$, applying Kolmogorov-Smirnov goodness-of-fit test, etc.).

\section{Data and results}

The phenomena analyzed in this paper and their corresponding data sets are described below.
Complete information about the outcome of our statistical analysis is provided in Tables 1-6.
Table 1 shows the results of the fitting of an untruncated power-law distribution.
Table 2 complements these results with the uncertainty of the parameters obtained from bootstrap.
Table 3 shows the results for a truncated log-normal, only for the data sets
for which this distribution yields a better fit than the untruncated power law.
Table 4 corresponds to the fitting of a truncated power law,
only when this distribution leads to a different fit than the untruncated case.
Table 5 gives the uncertainty obtained by bootstrap of the resulting parameters.
Table 6 summarizes the results with the preferred distribution for each data set.

Figures 1 and 2 display the empirical distributions together with the preferred fits in each case.
These figures play no role in the fitting procedure, and are shown only for the sake of illustration,
to have a visual and intuitive perspective.
The empirical distributions are properly normalized; the fits,
as defined in general over a smaller range, are not normalized but adapted to the normalization of the
empirical distributions (see the captions for detailed information).

\subsection{Earthquakes}

As mentioned in the introduction,
earthquake size distributions have been related to power laws since the 1930's
\cite{Utsu_GR}.
In 1944 Gutenberg and Richter
published their celebrated relation for the number of earthquakes
in terms of their magnitude \cite{Gutenberg_Richter}, which turns out to be an exponential relation
(some literature shows a certain confusion when mentioning ``power-law earthquake magnitudes'').
The reason for the exponential relation is that magnitude is a logarithmic measure of size, e.g.,
magnitude is the logarithm of energy \cite{Kanamori_rpp},
and when the Gutenberg-Richter relation is represented in terms of a more physical ``size''
(energy, seismic moment, rupture area, etc.)
a power law should be recovered \cite{Burridge_Knopoff}.
The Gutenberg-Richter relation, or law, has revealed as a very robust pattern of earthquake occurrence,
and it has been claimed that their power-law exponent turns out to be nearly universal,
in the sense that for many regions of the world and for different ranges of sizes, it takes values close to
$5/3\simeq1.67$ \cite{Godano_Pingue,Kagan_pageoph99,Kagan_gji02,Kagan_tectono10}.
However, other authors have claimed systematic variation of the power-law exponent
\cite{predict_b_nature},
which may be related to different tectonic stress regimes \cite{Schorlemmer}.

Some authors have argued that an untruncated power-law distribution
is problematic, as extrapolation of this relation to the largest earthquakes would lead
to an infinite release of energy in the long term, due to the divergence of the mean
value of power-law distributions when $\beta \le 2$
\cite{Knopoff_Kagan77,Serra_Corral}.
So, deviations from the Gutenberg-Richter power-law behavior should be expected for the largest earthquakes;
nevertheless, we will show that the untruncated power law is not such a bad model
for earthquake sizes.

We analyze the centroid moment tensor (CMT) catalog
\cite{Ekstrom2012},
which records (among other variables) the seismic moment of earthquakes across the globe
(so, $x$ is the seismic moment, measured in dyn cm).
The temporal period of our study goes from January 1, 1977 to August 31, 2017.
A refined analysis would have separated shallow, intermediate, and deep events
or, even better, different tectonic zones \cite{KaganBirdJackson};
however, we have considered the overall catalog.
In addition, we also analyze seismicity from Southern California (USA),
using the catalog of \citeA{YHS2012}, which covers the 30-year period 1981--2010, contains focal mechanisms,
and includes the moment magnitude $m$, which can be directly converted to seismic moment $x$ (in dyn cm)
by means of the formula \cite{Hanks_Kanamori}
$$
x=10^{1.5m +16.1}.
$$

The logarithmic-coefficient-of-variation test applied to the CMT catalog
indicates that an untruncated power-law distribution
is not preferred in front of a log-normal for a cut-off $a$ smaller that $2\cdot 10^{27}$ dyn cm
(corresponding to 7.5 in moment magnitude).
So, the power-law fitting procedure with $a$ restricted to be above this value leads to $\beta \simeq 2.1$,
for 2 orders of magnitude
(this is significantly different than the usual value $\beta\simeq 5/3$).
Nevertheless, this does not preclude that other distributions can fit the data even better
\cite{Serra_Corral}.
In any case, the log-normal is not one of this, and no log-normal fit is found.
The fit of a truncated power law restricted to upper cut-offs $b$ below the crossover value
$2\cdot 10^{27}$ dyn cm
leads this time to the usual value, namely, $\beta=1.655$, for 3 orders of magnitude
(with a lower cut-off corresponding to 5.3 in moment magnitude).
As the fitting ranges of both power laws (untruncated and truncated) almost overlap,
we can conclude that a double power-law distribution with a crossover at 7.6 in magnitude
is a satisfactory result; this is in agreement with previous literature
\cite{Pacheconature,Yoder2012}.
This double power law, in addition, avoids the problem of the divergence of the mean energy,
as the exponent of the power-law tail is $\beta_2> 2$.
On the other hand, the fitting of an untruncated power law to the whole range of the data
would have not lead to the rejection of it ($\beta\simeq 1.67$),
due to the small weight of the deviations at the tail in front of the rest of the distribution.
This case illustrates the convenience of using the test on the logarithmic coefficient of variation
in order to guide the fitting procedure based on maximum likelihood and goodness of fit.
For California we find an untruncated power law with $\beta=1.655$ for almost 6 orders of magnitude,
starting at $a=1.5 \cdot 10^{21}$ dyn cm (3.4 moment magnitude).
This constitutes the largest fitting range found in this article.
The truncated power law leads to very similar results.
Moreover,
the log-normal fit is not preferred in this range, as indicated by the computation of $cv_\ell$
and the corresponding test.

\subsection{Karst sinkholes and closed topographic depressions}

Sinkholes are ground depressions produced by subsidence due
in most cases
to karst, that is, by dissolution of soluble rocks by groundwater.
Sinkholes pose a hazard since they may form under buildings and infrastructure (such as roads, railways and pipelines), and are particularly dangerous if they collapse suddenly, causing even human casualties \cite{Galve2011,Brinkmann2013}.

Power-law frequency-size relations have recently been proposed for sinkholes, considering either their diameter
or their area as a measure of size \cite{Galve2011,Wall2014,Yizhaq2017}, and their exponents have been proposed to vary as new sinkholes develop, grow and coalesce \cite{Yizhaq2017}.
The frequency-size relation of topographic depressions in Florida shows a similar behavior \cite{Wall2014}.
These depressions are used as proxies to karst features, albeit not all of them are sinkholes \cite{Arthur2007}.
Notice that a logarithmic function between cumulative frequency and sinkhole diameter proposed for other sinkhole data sets \cite{Taheri2015,Gutierrez2016} is equivalent to a power law with unit exponent.

Sinkhole maps can be either delineated manually, or by automatically identifying topographic depressions in digital elevation models, which may lead to differences in the resulting inventories \cite{Wall2017}. Here we use
the Kentucky (USA) sinkhole database, mapped manually and probably the largest sinkhole data set available,
with over 100,000 sinkholes \cite{Paylor2003};
the database of Florida (USA) closed topographic depressions, based on automatic mapping, comprising more than 160,000 depressions \cite{FDEP2004};
and a compound data set of more than 1000 sinkholes next to the Dead Sea \cite{Yizhaq2017}.
The size is measured in terms of the area ($x=A$), in km$^2$. Sinkholes may have a minimum area on the order of 0.1--1 m$^2$, which is observed in the Dead Sea database and in other detailed studies \cite{Galve2009}. In contrast, for the Kentucky and Florida databases, the smallest reported areas are expected to be limited by the resolution of the maps used for their compilation. Meanwhile, the largest sinkhole size mapped may be affected by border effects, especially in small geographic areas.

Our results show that
the two largest data sets (Florida and Kentucky),
although not corresponding exactly to the same geological phenomenon,
lead to similar results.
An untruncated power-law distribution can be fit for 1.5 and 2
orders of magnitude, respectively, with exponent $\beta$ very close to $2.5$.
The $cv_\ell$ test confirms these results, in the sense that in the resulting range
the power-law fit is preferred in front of the log-normal.
However, as the log-normal holds for a much larger range,
this distribution is, overall, preferred,
with $\mu\simeq -10$ and $\sigma\simeq 2.5$--$2.8$
Results for the truncated power-law distribution are similar to the ones of the untruncated case,
with a somewhat smaller exponent ($\beta\simeq 2.3$) for Kentucky.
A spurious power law that appeared for the Florida data
in the range of very small areas has been disregarded.

The description of
the Dead-Sea data set in terms of the power law is poorer.
An untruncated power-law distribution is not rejected for the
last order of magnitude (being generous),
but with less than 50 data points,
and with a larger exponent ($\beta\simeq 2.8$).
The logarithmic coefficient of variation confirms this range,
for which the log-normal tail is not preferred;
nevertheless, as the log-normal leads to a much larger fitting range,
the latter is preferred
to describe the distribution globally.
In this case, though,
the log-normal needs and upper cut-off, $b\simeq x_{(N)}$
(otherwise, for $b\rightarrow \infty$, the log-normal is rejected).

On the other hand, note that if the areas $A$ are transformed into linear dimensions $L$
by means of the relation $L \propto A^{1/2}$,
Eq. (\ref{zetabeta}) with $z=2$ indicates than a range of $\beta$
from $2.5$ to $2.8$ transforms to the range $4$--$4.6$ for $\beta_L$.

\subsection{Wildfires}

Forest-fire models were one of the most popular topics in the field of self-organized
criticality;
however,
as far as we know,
simulation results from these toy models were not contrasted
with observational data until the work of \citeA{Malamud_science}.
Before the self-organized-criticality epoch
fire-size distributions had been plotted by \citeA{Minnich}.
\citeA{Malamud_science}
found that areas burned by wildland fires in different places of the USA and Australia
(including paleo-fire records)
follow (untruncated) power-law distributions with exponent $\beta$ ranging from 1.3 to 1.5.
Later, \citeA{Malamud_fires_pnas} studied 18 different ecoregions of the USA,
finding power-law distributions with exponents from 1.3 to 1.75.
Many other statistical studies have been performed,
see for instance the citations of \citeA{Malamud_fires_pnas}.
Interestingly, some authors have claimed that burned areas in other regions are better
described by the log-normal distribution \cite{Corral_fires,Hantson_Pueyo16}.

\citeA{Hantson_Pueyo16}
generated burned-area maps of high resolution (30 m)
from satellite (Landsat) imagery,
for some of the more important fire-occurrence regions in the world.
The resulting database comprised eight regions from different ecosystems and climates,
in all continents (except Anctartica).
The burned area $A$, in ha (1 ha= 0.01 km$^2$), was used as a measure of fire size ($x=A$).
After their careful statistical analysis,
the authors found that only two regions were compatible with the truncated power-law hypothesis
(in Canada and Kazakhstan), with exponents $\beta$ equal to 1.0 a 1.3, respectively.
Four other sites showed a log-normal behavior,
whereas the remaining two (in Angola and Australia),
where not compatible with any of the two statistical models.
Here
we select as representative sites those at Angola and Canada,
and we reanalyze their wildfire records.
The former site is associated to open woodland, with data gathered along
4 (non-consecutive) years,
whereas the Canada data corresponds to boreal forest monitorized for 14 years.

In contrast to the original results,
our analysis shows that
the wildfire data from Angola can be well represented
both by an untruncated power-law and by a log-normal distribution.
The untruncated power law holds
for burned areas greater than about 60 ha
(corresponding to more than 3.5 orders of magnitude),
with an exponent $\beta \simeq 1.8$.
In the comparison with the log-normal fit, the power law is not rejected in this range.
However, the log-normal is valid for a larger range, starting at about 10 ha,
covering 4.5 orders of magnitude and comprising more than 4 times
the number of points comprised by the power law.
The fit of the truncated power-law distribution does not lead to remarkable differences
with respect the untruncated power law.
Therefore, the log-normal is the preferred distribution.

In contrast, the Canada data yields a very short power-law tail,
with a larger exponent
and for a more limited range, starting at about $12,000$ ha
and containing very few data points.
We can disregard this power law as spurious and embrace the log-normal fit
which is valid for $x\ge 3.3$ ha up to the largest value (67,000 ha)
and covers a significant part of the data (77 \% of all data points).
In addition, and in agreement with the original work,
a truncated power law with $\beta\simeq 1.15$ gives a good fit between 1 and 12,000 ha,
covering 97 \% of all data points.
As the truncated power law and the log-normal are defined over different ranges
we have no way to decide between them, and we consider the two fits as valid.
That is, we can say that the truncated power law holds for small $x$ and
the log-normal for large $x$, but without a clear transition between them.

These results, together with those of many other previous researchers,
make it clear that, in contrast to earthquakes,
wildfires cannot be characterized by nearly universal power-law exponents,
as originally noticed in a qualitative way by \citeA{Minnich}.

\subsection{Tropical cyclones}

The term tropical cyclone applies to hurricanes and typhoons,
which are the same phenomenon, with the only difference that the former happens
in the North Atlantic and the Northeast Pacific and the latter in the Northwest Pacific.
But the term also comprises weaker systems, as tropical storms and tropical depressions.
The categorization is stablished by the maximum sustained wind speed,
which in tropical depressions is less than 34 knots,
and in tropical storms is between 34 and 64 knots
($1$ knot $ = 0.5144$ m s$^{-1}$);
otherwise, the tropical cyclone is considered a hurricane or a typhoon
%(in the Southern Hemisphere and in the Indian Ocean there are other names)
(or a severe tropical cyclone or a severe cyclonic storm, in other ocean basins)
\cite{Emanuel_book}.

\citeA{Emanuel_nature05} introduced the so-called power dissipation index (PDI)
as a rough estimation of the energy released by tropical cyclones in some ocean basin
during a whole tropical-cyclone season.
Later, the PDI was applied to estimate the energy of
individual tropical cyclones \cite{Corral_hurricanes}.
It is defined as
$$
PDI=\sum_{j=1}^K v_j^3 \Delta t,
$$
where the index $j$ denotes the $K$ different records of a tropical cyclone,
at different times, separated by intervals of $\Delta t=6$ hours,
and $v_j$ is the maximum sustained wind speed of record $j$.
We will use then the PDI as a measure of tropical cyclone ``size''
(that is, $x=PDI$) and
will convert the PDI units to m$^3$ s$^{-2}$,
although a rough conversion factor to Joules has been proposed
(PDI multiplied by $5 \cdot 10^6$ should yield energy in Joules)
\cite{Emanuel_power,Corral_agu}.
A truncated power-law distribution of PDI was previously proposed
\cite{Corral_hurricanes}.

Tropical-cyclone records analyzed here correspond to the
North Atlantic (NAtl) and the Northeastern Pacific (EPac) basins,
for the periods
1966--2016 and 1986--2016, respectively,
and are obtained from
the NOAA (USA) HURDAT2 data set
\cite{NOAA_hurdat_new}.
We also consider an aggregated data set joining the two basins (NAtl+EPac),
for the period 1986--2016.
A map view of the trajectories of the tropical cyclones in both basins indicates that it may make sense
to consider the two basins together, as a single unified mega-basin, in some sense.

In all three cases we find that an untruncated power law may fit
the tail of the distributions, with exponent $\beta$ around 4 or larger, with great uncertainty
(as shown by bootstrap).
Nevertheless, this power law is rather marginal,
as it extends for less than one order of magnitude;
in addition, the increase of the apparent value of the exponent with $a$
shows that we are not dealing with a genuine power law.
Notice that the power-law exponent of the joined data set does not fulfill the law of harmonic means
reported previously \cite{Navas_pre2},
because the fitting ranges of each data set are somewhat different.
Nevertheless,
the log-normal fit is not preferred for the tail.
Other works have fit a truncated gamma distribution
(which contains an exponential tail)
to this kind of data
\cite{Corral_agu,ftg}.
When fitting a truncated power law, the results are in agreement with
the original reference \cite{Corral_hurricanes},
with an exponent $\beta$ in the range $1.1$--$1.2$,
for about 2 orders of magnitude.
On the other side, a truncated log-normal distribution with $b\rightarrow \infty$
does not fit the data.
Finally, note that, from the figure and the conversion factor presented above,
the most extreme tropical cyclones (in terms of energy) release an energy around
$3 \times 10^{11}$ m$^3$/s$^2 \simeq 1.5 \times 10^{18} $ J.

\subsection{Rain}

Rainfall has been traditionally studied in terms of rain collected during a fixed time period
(one day, or one month) at a single site (i.e., a point measure in space).
Peters and co-authors
\cite{Peters_prl,Peters_pre,Peters_jh},
and previously \citeA{Andrade},
challenged that approach,
defining the rain event over some site
as a continuum of rain occurrence in time.
In the ideal case, one should be able to measure rain with high resolution in time,
for instance 1 min.
In this way, for a single site in the Baltic Sea, \citeA{Peters_prl} obtained
a power-law exponent $\beta\simeq 1.35$
for the total rain collected during rain events.
A subsequent analysis for different sites of the globe,
where rain was recorded using different instruments,
found a nearly universal exponent $\beta$ in the range 1.0--1.2.

Later, \citeA{Peters_dragon_kings} took a different view,
defining the rain event not along time but across space,
that is, they considered the instantaneous area of precipitation clusters.
This was obtained as in percolation theory, aggregating
nearest-neighbor precipitating pixels corresponding to one time slice.
The resulting areas $A$, in km$^2$ where measured (i.e., $x=A$).
The data used came from the
precipitation radar of the Tropical Rainfall Measuring Mission
satellite.
\citeA{Peters_dragon_kings} noticed a power-law-like regime for the areas,
with exponent $\beta\simeq 2.05$.
In fact, a similar result was claimed much earlier by \citeA{Lovejoy81}
for (radar) rain areas in the tropical Atlantic, stating power-law behavior with $\beta\simeq 1.82$.
Few years later \citeA{Lovejoy_mandelbrot85}
reported $\beta\simeq 1.75$ in the same context.

Our reanalysis of \citeA{Peters_dragon_kings}'s data shows that the fit of an untruncated power law distribution leads to poor results.
The power law can only be fit to the most extreme events, comprising less than 200 data points
(out of more than 5 million) and covering less than half an order of magnitude.
Moreover, the apparent exponent shows an increase with the cut-off value.
The truncated-power-law fit does not lead either to very positive results, despite the fact
the graphical representation of the distribution shows a decreasing linear trend in a log-log plot.
The reason of this failure may be due to the astronomically large number of clusters.
This means that the uncertainty associated with the empirical distribution is very tiny,
and the goodness-of-fit test can detect the smallest differences with respect a real (ideal)
power-law behavior.
The best truncated power-law obtained (the one having larger ratio $b/a$)
holds for less than one order of magnitude and comprises less than 1 \% of the clusters,
showing significant variation with the change of $a$ and $b$;
thus, we disregard the obtained power law as not relevant.

More recently,
\citeA{Traxl}
obtained the total volume of precipitated water in rain events defined over space and time.
They employed the Tropical Rainfall Measuring Mission records, calculated in space-time cells
with a resolution of 3 hours in time and $0.25$ degrees in (two-dimensional) space,
covering the globe from -50 to 50 degrees in latitude, for the time period 1998--2014.
Spatiotemporal rain clusters were defined via next-nearest neighboring ``occupied'' cells in space
and nearest neighbors ``occupied'' cells in time (with a total of 26 neighbors per cell).
Cell occupation was understood as occupation by extreme rain, i.e., with high rain rates,
above certain rain-rate thresholds.
These thresholds were defined locally, using the per-cell 90-th percentile of
rain rate (restricted to values above 0.1 mm/h).
The resulting thresholds ranged from 2 to 10 mm/h.
The volume of rain for each cluster, measured in km$^3$,
was calculated by integrating the rain rate over the spatiotemporal cluster.
In a formula,
$$
V=\int_{\mbox{cluster}} R dS dt
$$
(so, $x=V$),
where the integral is understood as summation over the cells
with the mentioned resolution, and $R$ is the rain rate
defined over space ($dS$) and time ($dt$).
We stress that $V$ is not the spatial volume of the cluster
but the volume of rain precipitated by the spatio-temporal cluster.
Notice that one can convert precipitated volume into energy
through the latent heat of condensation of water (approx. 2500 J/g);
thus, 1 km$^3$ of precipitation is roughly equivalent to $2.5 \times 10^{18}$ J
of latent-heat release.
Precipitation depth measured by \citeA{Peters_prl,Peters_Deluca}
can be transformed into energy per unit area by means of
1 mm $\simeq 2.5 \times 10^{6}$ J/m$^2$.

The resulting rain events were classified into two groups:
land events (with 90 \% or more of their cells corresponding to land)
and ocean events (defined in an analogous way);
the rest of clusters (2.5 \% of the total) were disregarded.
The authors obtained that a truncated gamma distribution
(called truncated power law by them) is what better fitted the ocean data set,
whereas the Weibull distribution (called there stretched exponential)
fitted both the land data set and the aggregated land+ocean data set.
The corresponding power-law exponent given by the gamma distribution
was $\beta\simeq 1.71$ (the Weibull distribution does not have a clear power-law regime
when its shape parameter goes to zero, which is what the authors found).

We re-analyze the data of \citeA{Traxl}, kindly updated by the authors for the 20-year period 1998-2017.
The land and ocean data sets were defined by 80 \% or more cells in there.
Our results for the untruncated power-law fit lead to very short power-law tails
(one order of magnitude, or less)
with values of $\beta$ rather large
(again, the bootstrap method shows large variability in these values,
and a dependence of $\beta$ on $a$ is also present).
The tail of the combined data set land+ocean
is constituted exclusively by ocean events,
and leads to the same results as the ocean data set.
The logarithmic-coefficient-of-variation test
indicates that the log-normal fit is not preferred for those short tails.

On the other side, for the ocean clusters,
a truncated power-law distribution with exponent $\beta=1.7$
holds for two orders of magnitude, comprising more than one million data points.
As far as we are aware,
this is one of the power-law distributions with more data ever found
\cite{Clauset}
using rigorous statistical procedures.
The other two files do not show any truncated power-law behavior,
due to the great amount of data
(taking smaller random subsamples one may find that truncated power laws are accepted,
as well as log-normal tails,
but that would mean disregarding the majority of data).
Regarding the overall shape of the probability density,
it is a remarkable fact that this
displays a great similarity with the probability density of the PDI of tropical-cyclones
\cite{Corral_hurricanes},
both shown in Fig. 1.
Note from there that the energy released by the most extreme rainfall events is of the order of $2 \times 10^3$ km $\simeq 5 \times 10^{21}$ J.
This large difference with respect tropical cyclones may be due
to the fact that in that case the energy referred to kinetic energy \cite{Emanuel_power}
whereas for rainfall we consider total latent heat released .

The reason of the difference between our results and those of the original reference
\cite{Traxl}
is that these authors were comparing different distributions
by means of likelihood-ratio tests,
but without testing the goodness-of-fit of any of the distributions individually.
So, they determine
a relative comparison about which distribution is preferred in comparison with the others,
but they do not provide absolute judgements.

\subsection{Impact fireballs}

Fireballs are exceptionally bright meteors produced by the impact of asteroids or comets on Earth's atmosphere and are usually called bolides if they explode midair. Large enough impact air bursts, even if the impactor does not reach the surface, can produce damage and casualties by a variety of mechanisms, especially by wind blast, thermal radiation and atmospheric shock wave overpressure \cite{Rumpf2017}, being the latter recently illustrated by the damaging 2013 Chelyabinsk impact \cite{Brown2013,Heimann2013,Popova2013}.

The distribution of energy of asteroid and comet impacts with Earth is usually reported as a power law \cite{Brown2013,Boslough2015}
composed by different data sets which span different parts of the energy range (from small meteors to fireballs to calculated impact rates from known near-Earth objects).

Here we use the fireball and bolide data provided
by the \citeA{CNEOS2018}, based on reports by US Government sensors \cite{Brown2002,Brown2013,Boslough2015}. The total impact energy ($x$) for each event is reported in kt (kilotons of TNT), extrapolated from the measured total optical radiated energy using an empirical fit \cite{Brown2002}. The largest fireball included there is Chelyabinsk (with a reported total energy of 440 kt), which was also independently located and characterized, even at regional or global distances, using the records of ground shaking produced by the shock wave \cite{Brown2013,Heimann2013} and atmospheric infrasound \cite{LePichon2013}.

The data set includes events since April 1988, albeit, as other authors \cite{Brown2002,Brown2013,Boslough2015} we use only the data since 1994, as earlier years are substantially incomplete. Until July 2018, the database contains 748 events. The percent of the Earth's surface covered by the sensors has varied over that period \cite{Brown2002}, and on average may be on the order of 80\% for events with total energy $>$ 1 kt \cite{Brown2013}.

Our statistical analysis shows that fireballs are well described by an untruncated power-law distribution
for about 3 orders of magnitude,
with exponent $\beta\simeq 2$, but with large uncertainty.
The truncated power law leads to practically the same results as the untruncated case.
The fit of the log-normal distribution is also not rejected over more or less the same range
as the power law; however,
the $cv_\ell-$test confirms that the power law is preferred in front of the log-normal
distribution over that range.

\section{Conclusions and Discussion}

We have revisited the important problem of power-law distributions
in the size of
geological and geophysical structures
and
of catastrophic phenomena of geoscience.
Due to the delicate issues involved with power-law fitting and
the variety of approaches in the literature,
we have undertaken a revision of some relevant examples of previously proposed power laws,
in order to treat all systems within a unified framework
and to facilitate the comparison between the results of different systems.

For that purpose,
we have extended an existing method for fitting and goodness-of-fit testing
of power-law-like distributions.
The important issue regarding these distributions is that the range over which a power-law distribution may fit the data is unknown.
The method,
previously proposed by
\citeA{Peters_Deluca} and
\citeA{Corral_Deluca},
uses the same tools as the so-famous procedure of \citeA{Clauset}
(maximum-likelihood estimation and the Kolmogorov-Smirnov test),
but is different in spirit,
allowing the fitting of both untruncated and truncated power laws
(generalization to other distributions is straightforward).

Three important improvements related with the cut-offs are included.
First,
for untruncated power laws the fitting range
is constrained by a complementary
likelihood-ratio test (based on the residual coefficient of variation) that compares a general power-law tail with a general log-normal tail
\cite{Malevergne_Sornette_umpu}.
Only lower cut-offs for which the power law is preferred in front of the log-normal are contemplated.
Second, the variation of the apparent power-law exponent with the cut-offs
is studied in order to rule out spurious power laws that appear when the number of data is low.
And
third, the uncertainty in the lower and upper cut-offs of the power-law distributions
is evaluated by bootstrap.

The results show a large diversity of outcomes.
The energy released by impact fireballs is well fitted by an untruncated power law;
nevertheless, as the size of the database is low,
this result could change if much larger databases become available.
An important issue here is the completeness of the records for all impact energies,
or at least the proper evaluation of incompleteness for the different energy ranges.

For the seismic moment of global earthquakes, we find two power-law regimes,
the first ranging up to about 7.6 in moment magnitude,
with the usual exponent $\beta=1.655$,
and the second power law defined above the previous value,
with exponent $\beta\simeq 2$.
This was previously proposed by some authors, due to geometric constraints of earthquake ruptures above that threshold magnitude, such as \citeA{Pacheconature} and \citeA{Yoder2012}.
In contrast, Southern-California seismicity shows a unique untruncated power-law range,
starting at around 3.4 in moment magnitude.
This value, which can be interpreted as a measure of the completeness threshold of the
analyzed catalog, may seem to be somewhat large,
but let us recall that our fitting procedure is rather demanding,
rejecting any power-law fit whose $p-$ value
is below 0.20.
Indeed, the value is coincident with the independent estimates of the detection threshold of the Southern California Seismic Network during the period considered \cite{Schorlemmer_Woessner}.
Despite the relatively large completeness threshold we find,
the resulting power law holds for almost 6 orders of magnitude (for seismic moment),
which is a rather impressive result.

Energy of tropical cyclones in two different ocean basins \cite{Corral_hurricanes}
is found to display truncated power-law behavior
over the central part of their distributions.
Here, problems at small energies are especially important due to incompleteness
of the records for small storms,
whereas the largest events (the most extreme hurricanes)
are strongly influenced by the boundary conditions imposed by
the finite size of the basins over which they develop, apart from other factors \cite{Chavas2016}.
The small size of the tropical-cyclone databases
does not allow to guarantee that the power-law behavior is maintained if much larger records
become available in the far future.

Rainfall clusters
\cite{Peters_dragon_kings,Traxl}
are in the opposite side of the spectrum regarding the number of data,
which is ``astronomical''.
Except in one case, we are unable to find meaningful power-law or log-normal fits.
The reason, naturally, is the millions of events comprising the databases,
which make almost impossible for goodness-of-fit test to accept
(i.e., not reject) any proposed distribution.
The remarkable exception is that of the total precipitation released
by spatio-temporal rainfall clusters over the oceans \cite{Traxl},
for which we find a truncated power law ranging only for two decades
but comprising more than one million data points.
We are not aware of a power-law distribution established with rigourous statistical protocols
containing more points.

On the other hand, sinkholes and topographic depressions
\cite{Paylor2003,FDEP2004,Yizhaq2017}
are better described by truncated log-normal distributions, down to minimum area thresholds probably controlled by the resolution of the databases used. Although power-law tails may fit narrow ranges of the largest areas reported, the log-normal distributions are preferred, as their fits cover a much larger range (including therefore a larger fraction of the entries contained in the catalogs).

The two examples of wildfires analyzed here
\cite{Hantson_Pueyo16}
are also well fitted by truncated log-normals over large ranges,
although for the Canada record this is not incompatible with a power law regime
(due in part to the low number of wildfires recorded there, both fits overlap over a significant range).

As this paper was pretended as an essentially statistical one,
an issue that we have not considered so far is the physical reason
behind power-law distributions.
Now we pay some attention to that before ending.
For earthquakes and wildfires it is recognized that one feasible mechanism
can be self-organized criticality (SOC) \cite{Bak_book,BTW87},
as both phenomena are characterized by an activity front
that propagates (somewhat fast) through a substrate,
in an avalanche-like manner.
The activation can be modelled as a branching process,
but these processes only yield power-law distributions
if they are precisely at their critical point \cite{Corral_FontClos}.
So, the self-organization mechanism ensures that criticality is achieved
``spontaneously'' in these systems, through a balance between driving and dissipation, and so the substrate has to be at the onset of instability (all the time).
But although the SOC mechanism is a plausible one for these systems
\cite{Bak_Tang,Sornette_Sornette,Ito_Matsuzaki},
it is a extraordinarily difficult task to establish that SOC is the responsible of
power-law size distributions in earthquakes, as there is a big gap
in the simplicity of SOC models \cite{OFC} and the enormous complexity of the real phenomenon.

Regarding wildfires,
we have already seen (as some previous authors)
that these seem to be better described by log-normals,
at least for some particular data sets.
We are not aware of modifications of SOC models that yield log-normal
distributions, although the non-applicability of the SOC mechanism
for wildfires in some cases
was discussed by \citeA{pueyo2010}.
Notice also that the sinkhole model proposed by \citeA{Yizhaq2017}
was claimed to lead to power-law distributions,
although no goodness-of-fit was performed there
and the results could be compatible also with the log-normal
(remember that our preferred model in that case was log-normal).
It is noteworthy that the log-normal distribution arises naturally for simple multiplicative random growth process, where the rate of growth is proportional to size by a random factor, as reviewed by \citeA{Mitz}. This can easily be seen just applying the classical central limit theorem to evolution of the logarithm of the size.

The fact that power-law distributions, similar to the Gutenberg-Richter law for earthquakes,
also seemed to characterize rainfall led some authors to speculate that
precipitation could be a sort of earthquake (avalanche) process in the sky
\cite{Peters_prl,Peters_pre,Peters_jh}. The metaphor is very appealing
but we are not aware of any physical model of rainfall production in terms of avalanches (for an alternative explanation, see \citeA{Dickman_rain}).
Nevertheless, indirect proof has been gathered by the identification of a
critical point in the transition to strong convection,
to which the atmosphere seems to be ``attracted'' \cite{Peters_np}.
Interestingly, tropical cyclones are related to rainfall
(they are a particular instance of it, an extreme case);
so, one can expect a connection between a SOC description of rainfall
and a SOC description of tropical cyclones,
although there are different options about which are the variables
that would support the SOC description
\cite{Corral_Elsner,Peters_dragon_kings}.
For rainfall, one may consider that the release of energy takes place
in the clouds, whereas for tropical cyclones one could locate the interaction
in the ocean-air interface.
Moreover, an important difference between tropical cyclones and earthquakes is that the former are
dynamical and thermodynamical organized structures
(one may talk about dissipative structures),
whereas earthquakes lack this sort of machinery.

Needless to say, nowadays it is well known that criticality and SOC
are not the only mechanisms able to generate power-law distributions
\cite{Sornette_critical_book,Mitz,Newman2004a,Simkin11}.
A particularly simple but interesting alternative
was proposed by \citeA{Reed}.
In the context of Zipf's law, different power-law-like models
were proposed much earlier than SOC
\cite{Simon,Miller_monkey},
like preferential growth or cumulative advantage (rich-get-richer)
\cite{Simon,Zanette_2005,Zanette_book,Cattuto},
which could make sense as well applied to growth processes in geophysics.
Another interesting option, if the power-law exponent is larger than 2,
are stochastic differential equations with additive and multiplicative noises
\cite{Penland,Penland2}.
However, other Zipf-like models are more difficult if not impossible to translate into geophysics
\cite{Miller_monkey,Tria,Corominas_dice,Ferrer_cancho_compression}.

Finally, several authors have used some limit theorems in probability theory to justify the origin of power-law distributions.
For instance, the generalized central-limit theorem
ensures that when a fixed number of (independent and identically distributed) random variables are added, the result converges to a L\'evy distribution
if some conditions (that invalidate the application of the classical central-limit theorem) are fulfilled \cite{Bouchaud_Georges}.
An analogous (but different) limit exists when the number of added random variables is not fixed but random, according to some prescribed distribution;
the limit distributions for a geometric-distributed number of terms (geometric stable distributions) turn out to be the Mittag-Leffler distributions
\cite{Gnedenko,Corral_jstat,Corral_csf}.
In the same way, if instead of summing one considers maximization,
then the extremal types theorem applies, leading to the generalized-extreme-value distribution;
and if one considers threshold exceedances and shifting,
then the Pickands-Balkema-de Haan theorem leads to the generalized Pareto distribution \cite{Coles}.
All these limit distributions contemplate power-law-like tails;
however, these tails are only obtained when the tails of the individual
distributions (those that are added, maximized, etc.) are of the same kind.
This fact excludes limit theorems as a genuine mechanism to generate power laws, as they ultimately deal with the trivial case of ``power-law-in, power-law-out'', in which a power-law-like distribution is obtained from the input of another, unexplained, power-law-like distribution.

\acknowledgments
We are very grateful to all researchers who have shared data with us:
Alfredo Hern\'andez,
Abigail Jim\'enez,
V\'{\i}ctor Navas-Portella,
Patricia Paredes,
Ole Peters,
Salvador Pueyo,
Dominik Traxl,
and
Hezi Yizhaq.
John Wall provided valuable comments on sinkhole databases
and Pedro Puig and Isabel Serra provided statistical wisdom. Insight provided by the Editor (C\'ecile Penland) and the anonymous reviewers is also appreciated.
We acknowledge
support from the Spanish Ministry of Science, Innovation and Universities (projects
FIS2015-71851-P, MAT2015-69777-REDT, Mar\'{\i}a de Maeztu Program for Units of Excellence in R\&D MDM-2014-0445, and Juan de la Cierva research contract FJCI-2016-29307 hold by A.G.).
The data used in this paper is not original and has been provided by the authors quoted
in the references and these acknowledgements.
%as well as 2014SGR-1307, from AGAUR.
%Enter acknowledgments, including your data availability statement, here.

%% ------------------------------------------------------------------------ %%
%% References and Citations
%%%%%%%%%%%%%%%%%%%%%%%%%%%%%%%%%%%%%%%%%%%%%%%
% BibTeX is preferred:
%

\newpage

 \begin{table}
 \caption{Results of the untruncated power-law fit.
The data sets contain $N$ events, the maximum value of the variable is $x_{(N)}$,
the number of data in the power-law range is $n$,
the lower cut-off of the fit is $a_{fit}$, the exponent is $\beta$, and the $p-$value is $p$.
The logarithmic range of the fit, or number of orders of magnitude
is $r=\log_{10}({x_{(N)}}/{a_{fit}})$.
The number of simulations is 1000, and
50 values of $a_{fit}$, equi-spaced in logarithmic scale,
are swept for each order of magnitude.
The approximated value of the transition from log-normal to power-law tail is also included,
and denoted $a_{cv}$.
When this value is marked by an asterisk
(in the right-most column)
it means that in principle
(when no restriction was applied on $a_{fit}$)
we found $a_{cv} > a_{fit}$,
then we enforced $a_{fit} > a_{cv}$.
In the rest of cases this was not required.
%In most cases
%$a_{cv} < a_{fit}$, so,
%the description in terms of a power-law tail
%is not improved by a log-normal tail
%(except when indicated by an asterisk in the last column).
EQ, TD, SH, TC, CA, and TP denote
earthquakes, topographic depressions, sinkholes, tropical cyclones,
rain cluster areas, and rain total precipitation, respectively.
}
 \centering
 \begin{tabular}{|l rrrr crcc | r|}
 \hline
data set & $N$ & units & $x_{(N)}$ & $a_{fit}$ & $r$ & $n$ & $\beta$ & $p$ & $a_{cv}$\\
 \hline
%EQ CMT global &     48,636 & dyn cm &  5.3$\cdot10^{29}$ &   3.2\cdot10^{24}$ &    5.22 &     12,224 &        1.666    &  0.21 & $^*$2$\cdot10^{27}$\\%  m_{min}=5.6              seismic_moment_cmt_eathq_1977-2017
%                             159
EQ CMT global &     48,636 & dyn cm & 5.3$\cdot10^{29}$
&    4.0$\cdot10^{27}$ &   2.12 &        97 &        2.087   &  0.33  & $^*$2$\cdot10^{27}$\\ %       trunc_cv_seismic_moment_cmt_eathq_1977-2017
EQ California
&    179,255 & dyn cm  &  1.1$\cdot10^{27}$ &  1.5$\cdot10^{21}$ &    5.89 &      2,883 &        1.655     &  0.30 & 3$\cdot10^{20}$\\%                         seismic_mom_YSH_2010.hash
TD Florida &    163,019 & km$^2$ & 45.7 &   1.0 &     1.66 &       476 &        2.472           &  0.39 & 0.8\\%                        Florida_sinkhole_areas_km2
SH Kentucky &    101,095 & km$^2$&   31.3 &   0.2 &     2.16 &       511 &        2.485           &  0.77 & 0.2\\%             Kentucky_sinkhole_areas_km2_CORRECTED
SH Dead Sea &      1,033 & km$^2$&  0.0041 &   0.00044 &      0.98 &        48 &        2.804           &  0.59 &  $0.0002$\\ %8e-5
%                                Dead_Sea_sinkholes
Fires Angola &     17,643 &  ha&  271,000 &   63 &     3.63 &      1,294 &        1.818           &  0.22 & 30\\%                                      fires_Angola
%Fires Canada &       408 & ha & 67,600 &   2,290 &     1.47 &        44 &        1.832           &  0.21 & $^*$10,000\\%                                      fires_Canada
Fires Canada &       408 & ha & 67,600  &   12,600 &     0.73 &        13 &        2.277       &  0.25 & $^*$10,000\\%                             trunc_cv_fires_Canada
%&     11,442 &   .340E+05 &   .316E+03 &     2.032 &       738 &        2.033         .038  &  .27 &\\%                                    fires_Colombia
%TC NAtl &       771 & $10^{10} $ m$^3$s$^{-2}$ & 25 &   4.8 &      0.73 &        81 &        2.880           &  0.45 & $^*$10\\%                         natl_pdi_new_data_git_lab
TC NAtl &       771 & $10^{10} $ m$^3$s$^{-2}$ & 25 &
   12.6  &   0.31 &        16 &        5.145      &  0.22 &  $^*$10\\%                trunc_cv_natl_pdi_new_data_git_lab
TC EPac &       594 & $10^{10} $ m$^3$s$^{-2}$ &   31 &   5.0&      0.79 &        76 &        3.275           &  0.35 & 5\\%                         epac_pdi_new_data_git_lab
%TC NAtl+EPac &      1,065 & $10^{10} $ m$^3$s$^{-2}$&   31 &   5.2 &      0.77 &       127 &        3.226           &  0.55 & $^*${8}\\%                    Natl_Epac_pdi_new_data_git_lab
TC NAtl+EPac &      1,065 & $10^{10} $ m$^3$s$^{-2}$&   31
 &   7.9 &   0.59 &        56 &        3.578          &  0.55
& $^*${8}\\%                    Natl_Epac_pdi_new_data_git_lab
Rain CA &   5,037,333 & km$^2$ &  238,000 &   100,000 &     0.38 &       165 &        6.479          &  0.50 & 100,000\\%
TP land &    6,385,195 & km$^3$ &  177 &   33 &      0.73 &       286 &        4.109           &  0.38  & 30\\%                           trunc1_traxl_sizes_land
%TP ocean &     10,372,063 & km$^3$&   2,080 &   182 &     1.06 &       661 &        3.186           &  0.21 & $^*$200\\%                         trunc10_traxl_sizes_ocean
TP ocean &     10,372,063 & km$^3$&   2,080
&   200 &     1.02 &       555 &        3.250           &  0.47 & $^*$200\\%
%land+ocean &    16,757,258 & km$^3$ &   2,080 &   182 &     1.06 &       661 &        3.186     &  0.22 & $^*$200\\%                     trunc1_traxl_sizes_land_ocean
TP land+ocean &    16,757,258 & km$^3$&   2,080 &   200 &     1.02 &       555 &        3.250           &  0.47 & $^*$200\\%
Fireballs &       748 & kt &   440 &   0.3 &     3.14 &       278 &        2.022          &  0.21 & 0.2\\%                                         Fireballs
% "Calculated total impact energy in kt (kilotons of TNT)"
\hline
%& \multicolumn{2}{l}{$^{a}$Footnote text here.}
 \end{tabular}
 \end{table}

\begin{table}
 \caption{
Application of the bootstrap procedure to the untruncated power-law fit.
The exponent $\beta$ is represented by the mean of all bootstrap results, $\bar \beta$,
and variability in $\beta$ by one standard deviation $\sigma_b$.
The $p-$value is also given in terms of the mean value $\bar p$.
The variability of lower cut-off $a$ is calculated by taking its logarithm,
calculating the mean and the standard deviation, and transforming back
taking the exponential of the results.
The three values reported are associated to the mean of the logarithm and to this plus/minus one standard deviation.
The number of bootstrap samples is 100.
}
 \centering
 \begin{tabular}{|l |rrr| c | c|}
 \hline
data set & & $a_{fit}$ &   & $\bar \beta \pm \sigma_b$ & $\bar p$ \\
 \hline
%%EQ CMT global & 1.4$\cdot 10^{24}$ & 2.0$\cdot 10^{25}$ & 2.9$\cdot 10^{26}$ & 1.75$\pm$0.18 & 0.31\\
EQ CMT global & 2.6$\cdot 10^{27}$ & 4.4$\cdot 10^{27}$ & 7.4$\cdot 10^{27}$ & 2.05$\pm$0.16 & 0.35\\
EQ California &2.1$\cdot 10^{21}$&7.2$\cdot 10^{21}$& 2.6$\cdot 10^{22}$& 1.65$\pm$0.04 & 0.29 \\
TD Florida &              0.52&1.0&1.9 &             2.50$\pm$0.16   &  0.33 \\%                        Florida_sinkhole_areas_km2
SH Kentucky &      0.07&0.2&0.5  &             2.47$\pm$0.24   &  0.33 \\%             Kentucky_sinkhole_areas_km2_CORRECTED
SH Dead Sea &     2.4$\cdot 10^{-4}$ & 4.5$\cdot 10^{-4}$ & 8$\cdot 10^{-4}$ &
                                                           2.89$\pm$0.66    &  0.32  \\ %8e-5
%                                Dead_Sea_sinkholes
Fires Angola &            40&120&370    &     1.79$\pm$0.07  &  0.29 \\%                                      fires_Angola
%Fires Canada &           820&2,900&9,900&    2.15$\pm$0.86    &  0.28 \\%                                      fires_Canada
Fires Canada &           12,000&15,300&19,500&    2.85$\pm$0.65    &  0.30 \\%                                      fires_Canada
%TC NAtl &              4.8 & 7.7 & 12.3 &
%                                                           4.54$\pm$2.08    &  0.34 \\%                         natl_pdi_new_data_git_lab
TC NAtl &              11.1 & 12.6 & 14.4 &
                                                           5.88$\pm$1.62    &  0.39 \\%                         natl_pdi_new_data_git_lab
TC EPac &             4.7 & 6.5 & 8.9 &
                                                           3.89$\pm$1.14    &  0.38 \\%                         epac_pdi_new_data_git_lab
%TC NAtl+EPac &5.0&7.5&11.1&3.78$\pm$0.83& 0.35\\
TC NAtl+EPac &8.0&10.8&14.7&4.54$\pm$1.61& 0.41\\
TP land &                   27&34&44 &               4.29$\pm$0.57           &  0.38  \\%                           trunc1_traxl_sizes_land
%TP ocean           &   170&240&345 &              3.43$\pm$0.36           &  0.35 \\%                         trunc10_traxl_sizes_ocean
TP ocean           &   190&270&380 &              3.47$\pm$0.31           &  0.40 \\%                         trunc10_traxl_sizes_ocean
%land+ocean &    160&240&350         &        3.42$\pm$0.33           &  0.39 \\%                     trunc1_traxl_sizes_land_ocean
TP land+ocean           &   190&270&380 &              3.47$\pm$0.31           &  0.40 \\%
Fireballs &         0.4&0.6&1.0      &        1.95$\pm$0.35           &  0.37 \\%                                         Fireballs
\hline
%& \multicolumn{2}{l}{$^{a}$Footnote text here.}
 \end{tabular}
 \end{table}

\begin{table}
 \caption{Results of the log-normal fit for the data sets for which this leads to better results
that the untruncated power law, in the sense that the fitting range is remarkably larger.
$\mu$ and $\sigma$ are the log-normal parameters;
note that other researchers prefer to report
$e^\mu$ and $e^\sigma$ instead \cite{Limpert_lognormal}.
The upper cut-off is fixed to $b^{-1}=0$,
except for the Dead-Sea sinkholes.
The number of simulations is 1000, and
50 values of $a_{fit}$, equi-spaced in logarithmic scale,
are swept for each order of magnitude.}
\centering
 \begin{tabular}{|l |rrrcrccc|}
\hline
data set & $N$ & $a_{fit}$ & $b_{fit}$ & $r$ & $n$ & $\mu$ & $\sigma$ & $p $ \\
\hline
TD Florida &    163,019 & 0.018 & $\infty$ & 3.40 & 35,167 & -9.85 & 2.84 & 0.59 \\
SH Kentucky &    101,095 & 0.0095 & $\infty$ &3.51  & 17,624 & -9.70 & 2.54 & 0.22\\
SH Dead Sea &      1,033  &   $1.3\cdot10^{-6}$ &   0.005 &  3.60 &       964 &       -10.4  &   1.64  &  0.23\\%
Fires Angola &     17,643  &    8.7 &   $\infty$ &  4.49 &      5,349 &      -13.2   &   4.96  &  0.32\\%
Fires Canada &      408 &     3.3 &  $\infty$ &  4.3 &       314 &  3.36&  3.29   &  0.45\\%
%Fires Canada  &
\hline
%& \multicolumn{2}{l}{$^{a}$Footnote text here.}
 \end{tabular}
 \end{table}

 \begin{table}
 \caption{Results of the truncated power-law fit,
only for the data sets which lead to power-law ranges clearly different than
the untruncated case.
%The data sets contain $N$ events,
The number of data in the truncated power-law range is $n$,
the lower cut-off of the fit is $a_{fit}$,
the upper cut-off is $b_{fit}$,
the exponent is $\beta$, and the $p-$value is $p$.
The number of orders of magnitude covered by the fit is
$r=\log_{10}({b_{fit}}/{a_{fit}})$ .
The number of simulations is 1000
(except for the case of total precipitation of ocean rain clusters, which is 100),
and
10 values of $a_{fit}$ are swept equi-spaced in logarithmic scale
for each order of magnitude.
%SH and TC denote sinkholes and tropical cyclones, respectively.
}
 \centering
 \begin{tabular}{|l rrr c rcc |}
 \hline
data set & $N$ & $a_{fit}$ & $b_{fit}$ & $r$ & $n$ & $\beta$ & $p$ \\
 \hline
EQ CMT global &     48,636
%&     48477
&    1.3$\cdot10^{24}$ &   2$\cdot10^{27}$ &  3.20 &     22,061 &        1.655        &  0.46\\%      trunc2_cv_seismic_moment_cmt_eathq_1977-2017
%?? Florida &    142981 &   .457E+02 &   .100E+01 &   .501E+02 &  1.7 &       476 &        2.443         .072  &  .43\\%                  Florida_sinkhole_areas_km2_trunc
%SH Kentucky &    101095 &   .313E+02 &   .794E-01 &   .316E+02 &  2.6 &      1757 &        2.285         .032  &  .26\\%             Kentucky_sinkhole_areas_km2_CORRECTED
SH Dead Sea &      1,033 &     $2.5\cdot10^{-7}$ &   $1.3\cdot10^{-5}$ &  1.7 &       323 &         0.391           &  0.24\\%                                Dead_Sea_sinkholes
%Fires Angola &     17,643 &    .631E+02 &   .316E+06 &  3.7 &      1294 &        1.813         .023  &  .26\\%                                      fires_Angola
Fires Canada &       408 &     1.0 &   12,600 &  4.1 &       395 &        1.153           &  0.23\\%                                      fires_Canada
TC NAtl &       771 &     0.16 &   10 &  1.8 &       534 &        1.224           &  0.27\\%                         natl_pdi_new_data_git_lab
TC EPac &       594 &      0.08 &   10 &  2.1 &       511 &        1.085           &  0.37\\%                         epac_pdi_new_data_git_lab
TC NAtl+EPac &      1,065 &     0.16 &  10 &  1.8 &       777 &        1.156           &  0.26\\%                    Natl_Epac_pdi_new_data_git_lab
Rain CA &   5,037,333 &   6,310 &   50,100 &   0.9 &     15,817 &        1.711    &  0.32\\%                                      col5_WPAC_v5
TP ocean  &  10,372,063 &      0.126 &   12.6 &  2.0 &   1,273,883 &        1.721           &  0.66\\%
%Fireballs &       748 &   .440E+03 &   .501E+00 &   .398E+03 &  2.9 &       167 &        2.015         .083  &  .30\\%                                         Fireballs
\hline
%& \multicolumn{2}{l}{$^{a}$Footnote text here.}
 \end{tabular}
 \end{table}

\begin{table}
 \caption{
Application of the bootstrap procedure to the truncated power-law fit.
Only cases which lead to meaningful power laws are included.
The exponent $\beta$ is represented by the mean of all bootstrap results, $\bar \beta$,
and variability in $\beta$ by one standard deviation $\sigma_b$.
The $p-$value is also given in terms of the mean value $\bar p$.
The variability of both cut-offs $a$ and $b$ is calculated as in the untruncated case
by the mean plus/minus one standard deviation of their logarithms.
The number of bootstrap samples used is 100,
except for the total precipitation at the ocean,
which is 30.
}
 \centering
 \begin{tabular}{| l | rrr | rrr | c | c|}
 \hline
data set &  & $a_{fit}$ &  &  & $b_{fit}$ & & $\bar \beta \pm \sigma_b$ &  $\bar p$ \\
 \hline
EQ CMT global & $1.2\cdot10^{24}$&$3.2\cdot10^{24}$& 8.9$\cdot10^{24}$ &
                     $1.0\cdot10^{27}$&$1.6\cdot10^{27}$& $2.6\cdot10^{27}$  &  $1.66\pm0.02$& 0.34 \\
%SH Dead Sea  & $9 \cdot 10^{-8}$ & $3\cdot 10^{-6}$ & $ 10^{-4}$ & $3\cdot 10^{-6}$ & $7\cdot 10^{-5}$ & $2\cdot 10^{-3}$ & $1.10\pm 1.08$ & 0.31\\
Fires Canada & 1.5 & 7.1 & 33 & $9,400$ & $27,000$ & $78,000$ & $1.21 \pm 0.08$ & 0.29\\
TC NAtl & 0.06 &0.14& 0.35 & 2.1 & 6.1 & 17 & $1.16\pm 0.25$ & 0.31 \\
TC EPac & 0.06 & 0.10 & 0.17 & 3.8 & 7.9 & 17 & $1.08\pm0.12$ & 0.30\\
TC NAtl+EPac &0.06 & 0.14 &0.30 &2.0& 6.4 & 20 & $1.08\pm 0.26$ & 0.32 \\
TP ocean &0.08&0.27&0.91& 2.5&6.3&16& $1.73\pm0.03$ & 0.35 \\
\hline
%& \multicolumn{2}{l}{$^{a}$Footnote text here.}
 \end{tabular}
 \end{table}

\begin{table}
 \caption{Summary of results showing the preferred distribution for each data set.
For the Canadian wildfires, there is a truncated power-law regime starting at the minimum $x$ followed by a truncated log-normal covering
%p to the largest $x$,
up to the largest $x$,
without a clear transition between both distributions.}
 \centering
 \begin{tabular}{|l l|}
 \hline
data set & prefered distribution\\
 \hline
EQ CMT global & double power law \\
EQ California & untruncated power law \\
TD Florida &    truncated log-normal \\
SH Kentucky &     truncated log-normal \\
SH Dead Sea &       truncated log-normal (with $b^{-1}\ne 0$) \\
Fires Angola &      truncated log-normal \\
Fires Canada &       truncated power law  +  truncated log-normal\\
TC NAtl &       truncated power law \\
TC EPac &      truncated power law \\
TC NAtl+EPac &    truncated power law \\
Rain CA &   none \\
TP land &   none \\
TP ocean & truncated power law\\
TP land+ocean &    none \\
Fireballs &  untruncated power law \\
\hline
%& \multicolumn{2}{l}{$^{a}$Footnote text here.}
 \end{tabular}
 \end{table}

\newpage

\begin{figure}[h]
\centering
\makebox[\textwidth][c]{\includegraphics[width=18cm]{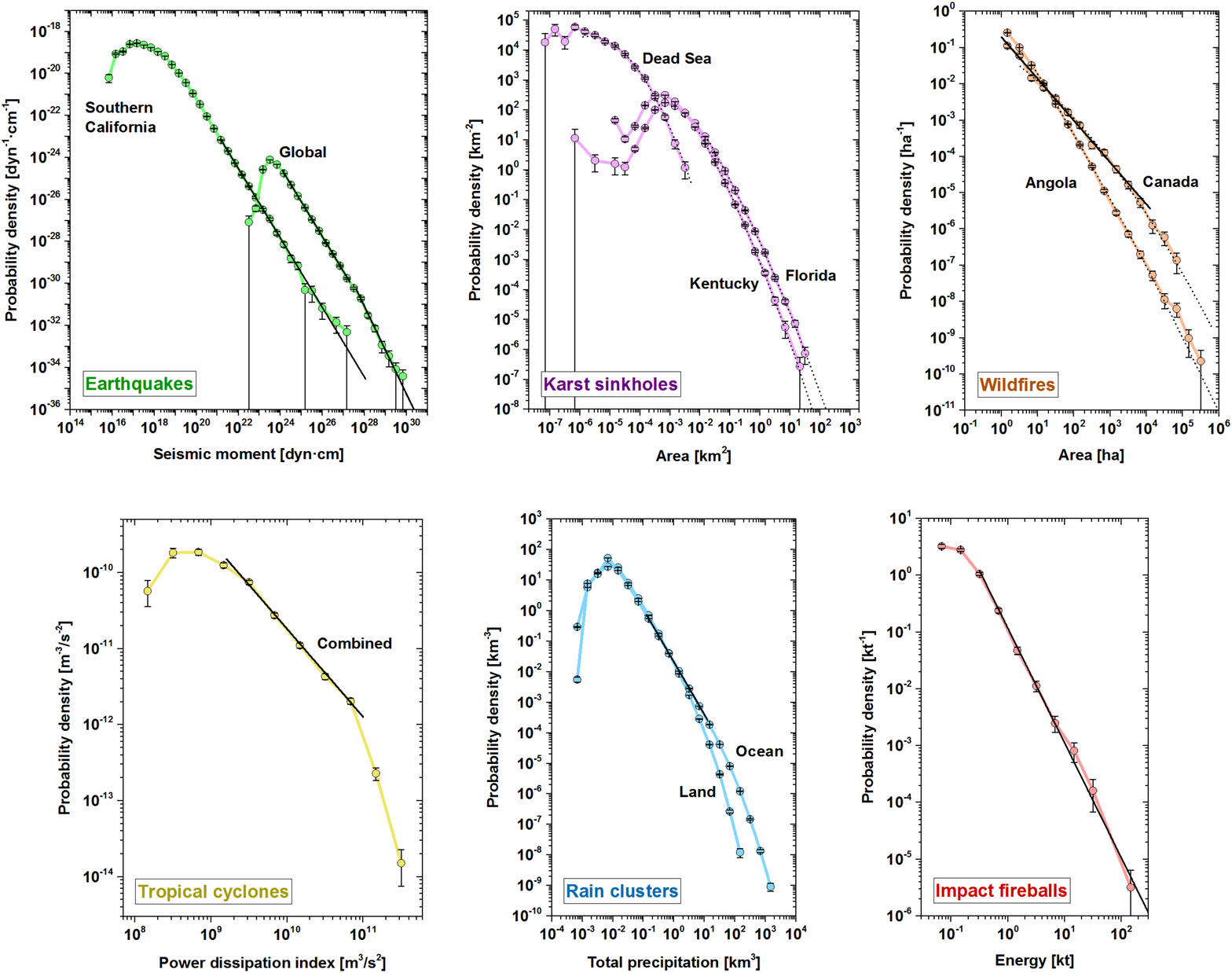}}%

\caption{
Estimation of empirical probability densities of size
together with the preferred fits for some of the data sets analyzed in this study.
Solid lines indicate power-law fits and dashed lines log-normal fits.
The theoretical distributions are rescaled as $n f(x)/N$ in order to properly fit
the representation of the empirical distributions, which are normalized over a different range.
The uncertainty of the empirical density represents one standard deviation
\cite{Corral_Deluca}.
}
% \label{figone}
  \end{figure}

\begin{figure}[h]
\centering
\makebox[\textwidth][c]{\includegraphics[width=18cm]{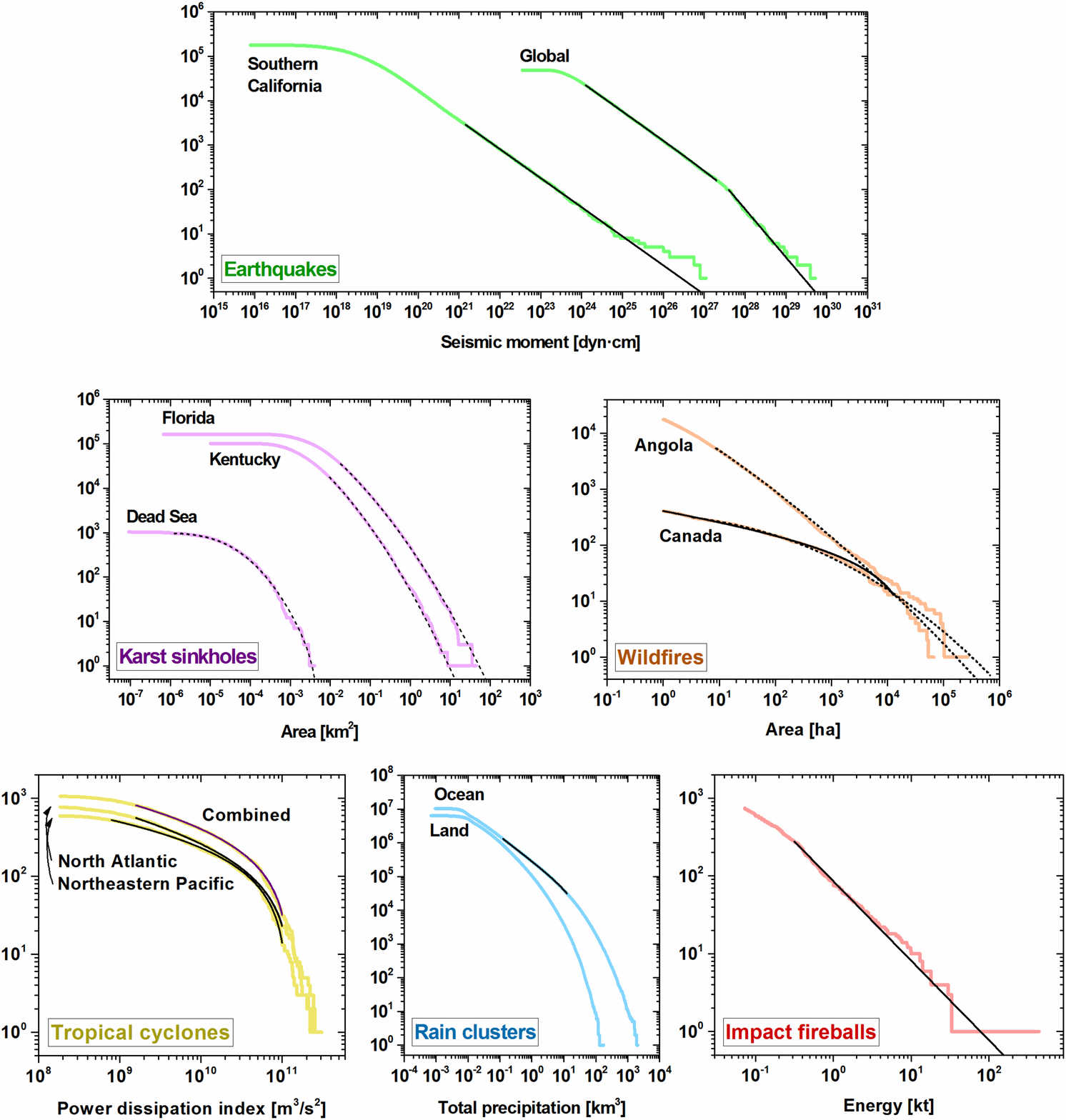}}%
\caption{
Number of structures or events with ``size'' larger or equal than the value of $x$
pointed in the axis,
together with the preferred fits for all data sets analyzed in this study
(except for the area of rain clusters and the total precipitation by clusters over land+ocean, which lead to no fit;
the total precipitation over land is included for visual comparison with the ocean case).
Solid lines indicate power-law fits, dashed lines log-normal fits.
Note that truncated power laws can be far from linear in the log-log plots,
especially for tropical cyclones.
In order to properly fit
the empirical results
the fits need to be rescaled and shifted as
$n S(x) + N S_{emp} (b)$,
with $N S_{emp}(b)$ the number of data with size at $b$ or larger.
}
% \label{figone}
  \end{figure}

\newpage
\bibliography{biblio}

\end{document}